# Enhancing Presence, Deepening Fan Intensity:

## How Presence in Immersive Video Shapes Psychological Closeness to Performers

**TOIDA Koichi[1], HIRANUMA Hideto[2], MIURA Shimpei[2], YAMAMOTO Norihiro[1], KOBAYASHI Yuki[1], and MEGURO Shingo[2]**

[1] MESON, Inc.
[2] Hakuhodo DY Holdings Inc., Marketing Technology Center.

Corresponding author: TOIDA Koichi (e-mail: koichi.toida@meson.tokyo)

**ABSTRACT** Immersive video is a video format that differs from conventional flat 2D video in that it is experienced as 180° stereoscopic video on a head-mounted display, thereby eliciting bodily and spatial subjective experience. Previous studies have shown that viewing distance and interpersonal distance affect Presence; however, it remains insufficiently understood how differences in Presence are related to psychological closeness to content. In the present study, we examined whether differences in Presence could increase immersive video viewers' psychological closeness to performers within the content. This psychological closeness was operationally defined as fan intensity in the present study. Specifically, a live performance by a Japanese idol group was recorded as 180° immersive video, and a high-Presence condition (1,200 mm) and a low-Presence condition (7,600 mm) were established by manipulating the filming distance. Twenty-four participants with different levels of prior involvement, comprising Avid fans and Casual fans, experienced videos from both conditions in a counterbalanced within-participants design. Fan intensity was measured before and after the experience as the perceived psychological overlap between the self and the performers. The results showed that, compared with the low-Presence condition, the high-Presence condition significantly increased all Presence-related measures except the Slater–Usoh–Steed questionnaire, with the largest condition differences observed for Possible Actions, Social Presence, and Observability. Moreover, a mixed analysis of variance on changes in fan intensity revealed a significant main effect of Presence condition, indicating that the high-Presence video produced a greater increase in fan intensity than the low-Presence video. These findings suggest that filming distance in immersive video is not merely a factor that determines angle of view or composition, but a design variable that can enhance Presence and deepen fan intensity.

# 1 INTRODUCTION

## 1.1 Background

Conventionally, content experiences have been positioned within a trade-off between a sense of presence and cost. In order to obtain an experience accompanied by a strong sense of presence and bodily participation, audiences have typically had to travel to the actual site and accept temporal, financial, and physical burdens. Conversely, when these costs are reduced, audiences have generally had to rely on media such as text, photographs, or flat video, thereby accepting a certain sacrifice in experiential vividness. In other words, the quality of an experience and the cost required to access it have long been implicitly regarded as opposing conditions.

In contrast, extended reality (XR) is a media form that can partially decouple and reconstruct the spatial and bodily sense of presence traditionally regarded as inherent to on-site experience from physical travel and on-site attendance. Accordingly, XR may be positioned as a media infrastructure that reorganises the trade-off between experiential quality and experiential cost, thereby enabling a third option: an experience that provides a strong sense of presence at a relatively low cost (Slater & Sanchez-Vives, 2016). In this sense, XR is not merely a collection of display technologies, but a media form that redefines the very mode of access to experience. Furthermore, immersive XR experiences have been shown not only to evoke the sense of being situated within a space, but also to affect socio-psychological responses such as self–other overlap, attitudes towards a target, emotional empathy, and helping behaviour through perspective-taking and embodied subjective experience (Ahn *et al.*, 2013; Herrera *et al.*, 2018; Martingano *et al.*, 2021; Pimentel *et al.*, 2021). In addition, in the context of content reception, preliminary findings have suggested that immersive XR experiences may influence understanding of the target, empathy, and behavioural intention (Kobayashi & Toida, 2026). Together, these findings suggest that XR may affect not only perceptual presence but also variables related to psychological relations with a target.

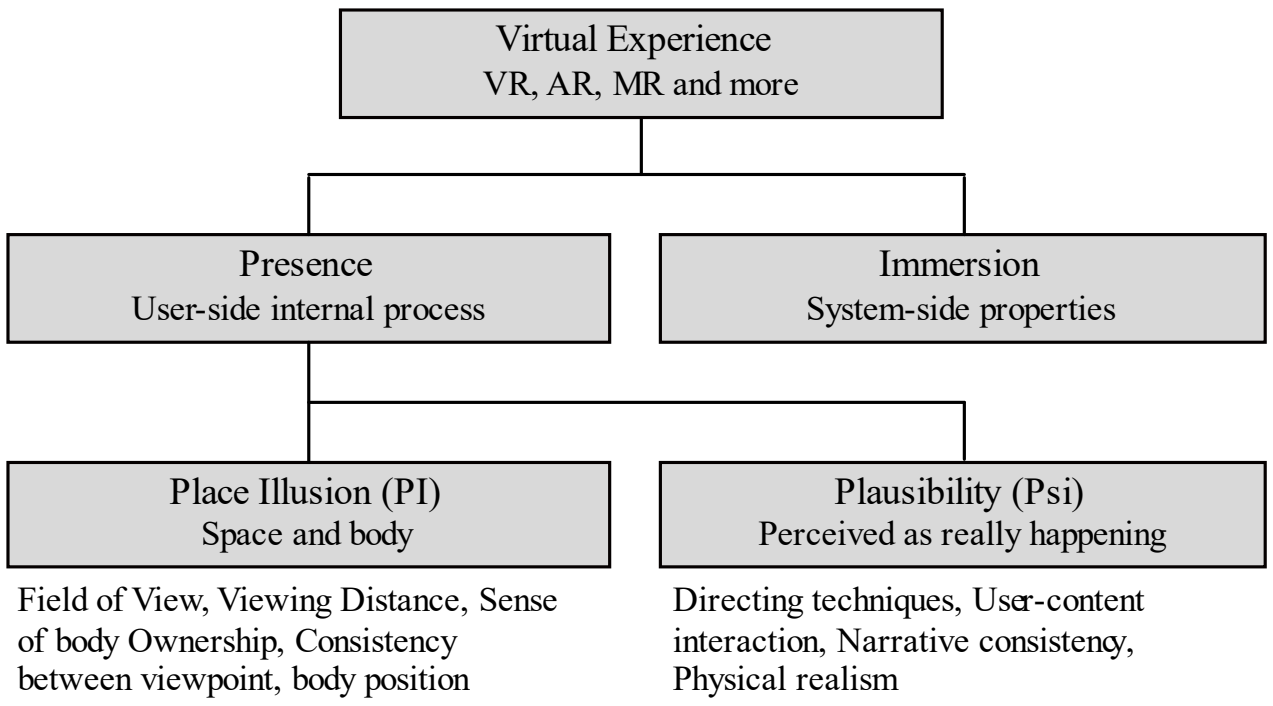

**FIGURE 1** Constituent elements of Presence and associated variables (based on Slater, 2009).

Immersive video experienced through XR devices occupies a distinctive position in this context. It is a video format that can elicit bodily and spatial subjective experience through 180° stereoscopic video presentation and spatial audio via a head-mounted display. It differs both from conventional flat 2D video viewing and from experiences in immersive virtual environments in which a two-dimensional video is presented with bodily and spatial qualities. That is, immersive video has a dual character: it is video, yet it is perceived as an experience. Moreover, in the production process of live-action immersive video, viewpoint position, filming distance, angle of view, and acoustic conditions can be treated as design variables. Thus, this format can be positioned as an experimentally manipulable stimulus that preserves live-action realism while eliciting bodily and spatial subjective experience.

To date, studies of virtual reality and immersive journalism have treated subjective experiences such as being-there, realism, and a sense of presence as primary evaluative dimensions (Schuemie *et al.*, 2001; Sundar *et al.*, 2017). However, it remains insufficiently understood how immersive video experiences alter attitude formation and relational evaluation towards a target through the viewer's perception, cognition, and affect, and through what psychological mechanisms such changes arise (Slater, 2009; Skarbez *et al.*, 2017).

## 1.2 Theoretical Framework of Presence

A central concept for theoretically understanding immersive experiences is *Presence* (Sheridan, 1992; Steuer, 1992). Experiences in virtual environments should not be discussed under the undifferentiated label of "immersion"; rather, they need to be understood by distinguishing between *Immersion* as a system-side property and Presence as a psychological state established on the user side (Fig. 1). Previous studies have defined the physical and technical conditions provided by a device, such as display resolution, field of view, stereoscopy, and tracking accuracy, as Immersion, whereas the subjective sense of "being there" that arises in users under those conditions has been defined as Presence (Sheridan, 1992; Steuer, 1992; Slater & Wilbur, 1997; Slater, 2009). Accordingly, subjective presence in virtual experience cannot be reduced simply to the richness of technical stimulation. Rather, Presence should be understood as a psychological state that is grounded in Immersion but constructed through the user's perception and cognition (Slater, 2009). This framework is also effective for evaluating immersive video experiences. What is critical is not the device or video format itself, but what kind of Presence is established in users through them.

Furthermore, Slater (2009) discussed Presence not as a single homogeneous concept, but as comprising at least two subcomponents (Fig. 1). The first is *Place Illusion*, a spatial and bodily illusion in which users feel that they themselves exist within the presented space. Place Illusion is a subjective phenomenon established on the user side and is related to self-location and bodily presence within the space. The second is *Plausibility Illusion*, which is based on the sense that the presented events or environment are coherent and that "what is happening there" is plausible. Plausibility Illusion primarily depends on the structure of the experienced content, the coherence of events, and the validity of narrative and staging. Although these two components are interrelated, they can, in principle, be manipulated through different design variables (Slater *et al.*, 2022; Skarbez *et al.*, 2017; Skarbez *et al.*, 2021).

The present study explicitly relies on this theoretical framework. In particular, it focuses on the Place Illusion aspect of Presence, which can be experimentally designed and manipulated in immersive video. However, filming distance is not the only design variable related to Place Illusion. Field of view, viewpoint height, gaze direction, body ownership, and the congruence between viewpoint and bodily position may also affect Presence. Among these factors, the present study focuses on filming distance from the performers within the content, because this variable can be directly manipulated during the production of live-action immersive video and can also be connected to the theoretical framework of Proxemics. By manipulating filming distance, we examined whether the extent to which participants felt that they were "there" could be altered, and how the resulting differences in Presence were related to changes in fan intensity.

## 1.3 Place Illusion, Proxemics, and Psychological Closeness

Place Illusion concerns the sense that experiencers locate themselves within the presented space. This sense is thought to be affected by geometrical conditions such as viewpoint height, gaze direction, and distance from the target (Slater, 2009). In this regard, the question of how people perceive distance from others or from objects, and what psychological meaning such distance has for interpersonal relations, has been systematically discussed within the framework of Proxemics (Hall, 1966).

In Proxemics, the distance between the self and others is generally divided into four zones: Intimate distance, Personal distance, Social distance, and Public distance (Fig. 2). These distance categories are understood not merely as physical distances, but as carrying psychological meanings related to interpersonal relations, potential interaction, intimacy, and discomfort associated with intrusion (Hall, 1966; Argyle & Dean, 1965; Patterson, 1976). Thus, distance from a target is not only a spatial arrangement but also a psychological variable that constitutes a relation.

Findings from Proxemics have been applied not only to interpersonal distance in face-to-face situations, but also to the understanding of pseudo-interpersonal distance in virtual environments. Studies of virtual reality and 360° video have reported that the distance between the camera or viewpoint position and an avatar may function as a pseudo-interpersonal distance between the experiencer and the avatar (Bailenson *et al.*, 2003; Yee *et al.*, 2007). Probst, Rothe, and Hussmann (2021) applied proxemic distance categories to camera-distance design in cinematic virtual reality, showing that differences in camera distance systematically changed evaluations related to comfort, naturalness, Presence, and psychological distance from an avatar. Specifically, within a virtual reality environment, the 100 cm condition, corresponding to Personal distance, was rated as the most natural and comfortable, and was also associated with higher ratings of Presence and of the sense of receiving personal guidance from the avatar. By contrast, in the 45 cm condition, corresponding to Intimate distance, the avatar was perceived as intrusive, and stress and discomfort increased. The 250 cm condition, corresponding to Social distance, was generally acceptable, whereas in the 500 cm condition, corresponding to Public distance, Presence was lowest and psychological distance from the avatar was rated as greatest. These findings indicate that distance design in relation to an avatar in virtual environments can affect comfort, naturalness, social presence, and psychological distance from the avatar. To illustrate this relationship schematically, Fig. 2 superimposes predicted curves for Presence and discomfort associated with intrusion onto the proxemic zones.

However, insufficient evidence has yet accumulated regarding how differences in Place Illusion produced by viewing distance, or by filming distance as an experimental manipulation, are related to attitudes towards a target and evaluations of the relation to that target. Many existing studies have treated Presence itself, or subjective evaluations such as a sense of presence, comfort, and naturalness, as dependent variables, and have not sufficiently addressed changes in psychological relations with a target (Cummings & Bailenson, 2016; Skarbez *et al.*, 2017). Furthermore, even when attitudinal change is observed, it is not easy to distinguish whether that change derives from enhanced Presence or from a mere exposure effect caused simply by increased opportunities for contact (Zajonc, 1968).

This issue becomes particularly important when the content target is a person or group. Such targets are not merely informational stimuli; rather, they are objects with which recipients may form relations composed of multiple psychological elements, including liking, familiarity, identification, intention to support, and intention for continued contact. Such one-sided yet intimate relations have also been discussed in research on parasocial interaction and identification (Horton & Wohl, 1956; Cohen, 2001). Therefore, in order to examine the effects of immersive video on content targets, it is necessary to capture not only perceptual evaluation, but also changes in the psychological relation with the target itself.

Accordingly, the present study focuses on *fan intensity* as an attitudinal variable that can be examined within an experimental-psychological framework. In this study, fan intensity is not the name of an established standard scale, but an analytical concept that represents the intra-individual level of involvement with idols as content targets. It should be understood as a composite attitudinal

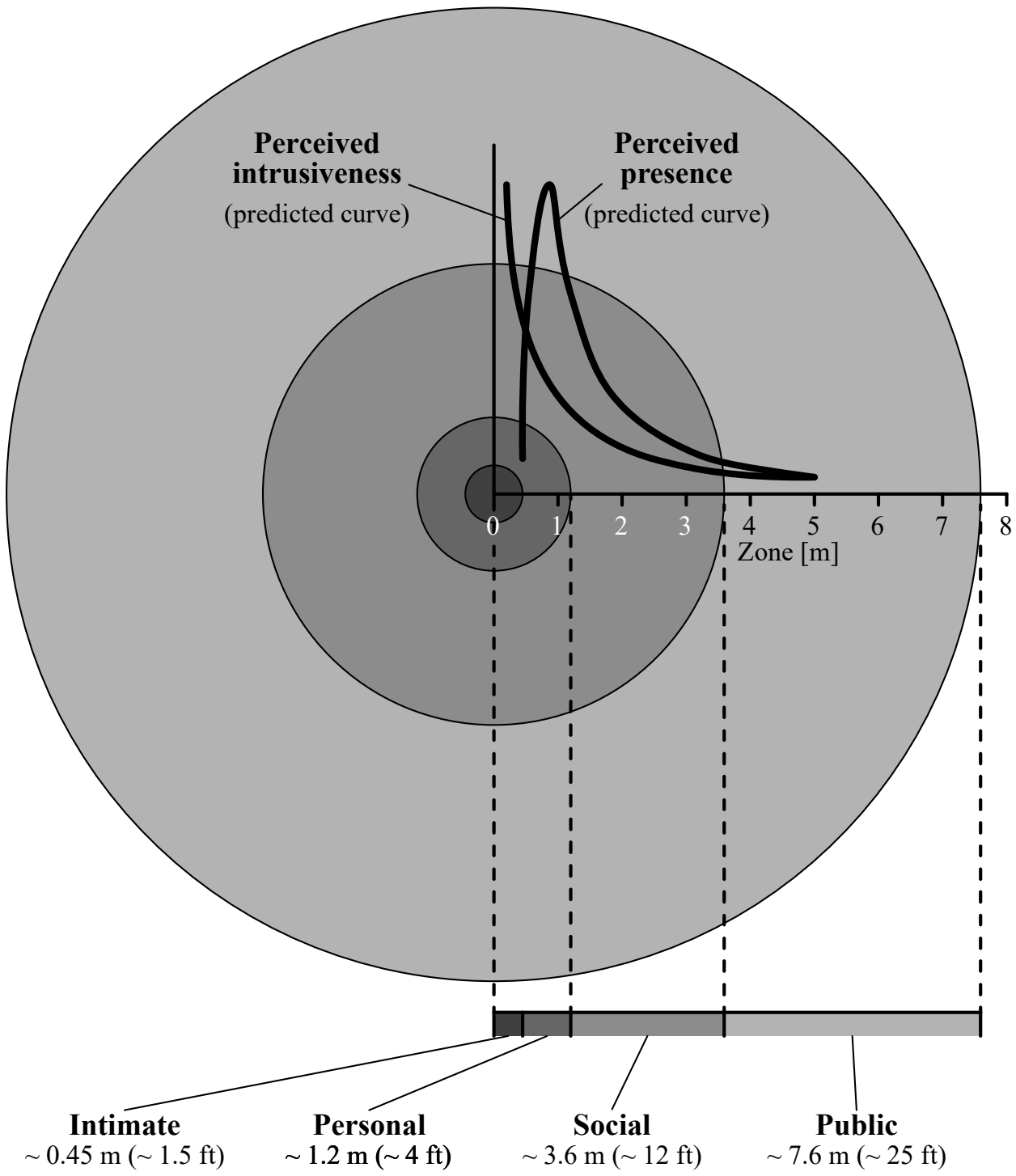

**FIGURE 2** Four zone categories based on proxemics (Hall, 1966), and predicted curves for perceived presence and discomfort associated with interpersonal intrusion across distance (Probst *et al.*, 2021)

level that cannot be reduced to mere liking, but includes familiarity, identification, intention to support, intention for continued contact, and psychological closeness in which the target is felt to be close to the self. Nevertheless, fan intensity is not a single, established psychological construct, and fandom research has also pointed to the multidimensionality of fan-related attitudes and behaviours, as well as the difficulty of defining them operationally (Hills, 2002). In light of this methodological constraint, the present study defines fan intensity in a limited sense, not as fan behaviour as a whole, but as psychological closeness to the content target. In other words, this study examines the extent to which recipients position the content target as close to the self through immersive video experience.

The present study uses an idol group as a target for which relation formation is likely to occur explicitly. Idol content was selected because relations with recipients tend to be formed explicitly and progressively, and psychological elements such as familiarity, intention to support, and intention for continued contact may change within a relatively short period of time. It is therefore suitable as a case for examining whether differences in Place Illusion in immersive video are related to fan intensity towards the content target.

## 1.4 The Present Study

On the basis of the foregoing discussion, at least three issues remain to be addressed in research on immersive video. First, the effects of immersive video should not be described merely as those of a "highly immersive video experience"; rather, it is necessary to clarify, on the basis of Presence theory, which subcomponents of Presence are associated with which psychological outcomes. Second, Place Illusion should be manipulated through experimentally designable variables, such as filming distance, and it should be examined whether the resulting differences are related to fan intensity towards the performers within the content. Third, this examination should be conducted not in terms of information comprehension or content evaluation, but in the context of fan psychology, where the formation of a relation with the target itself is central.

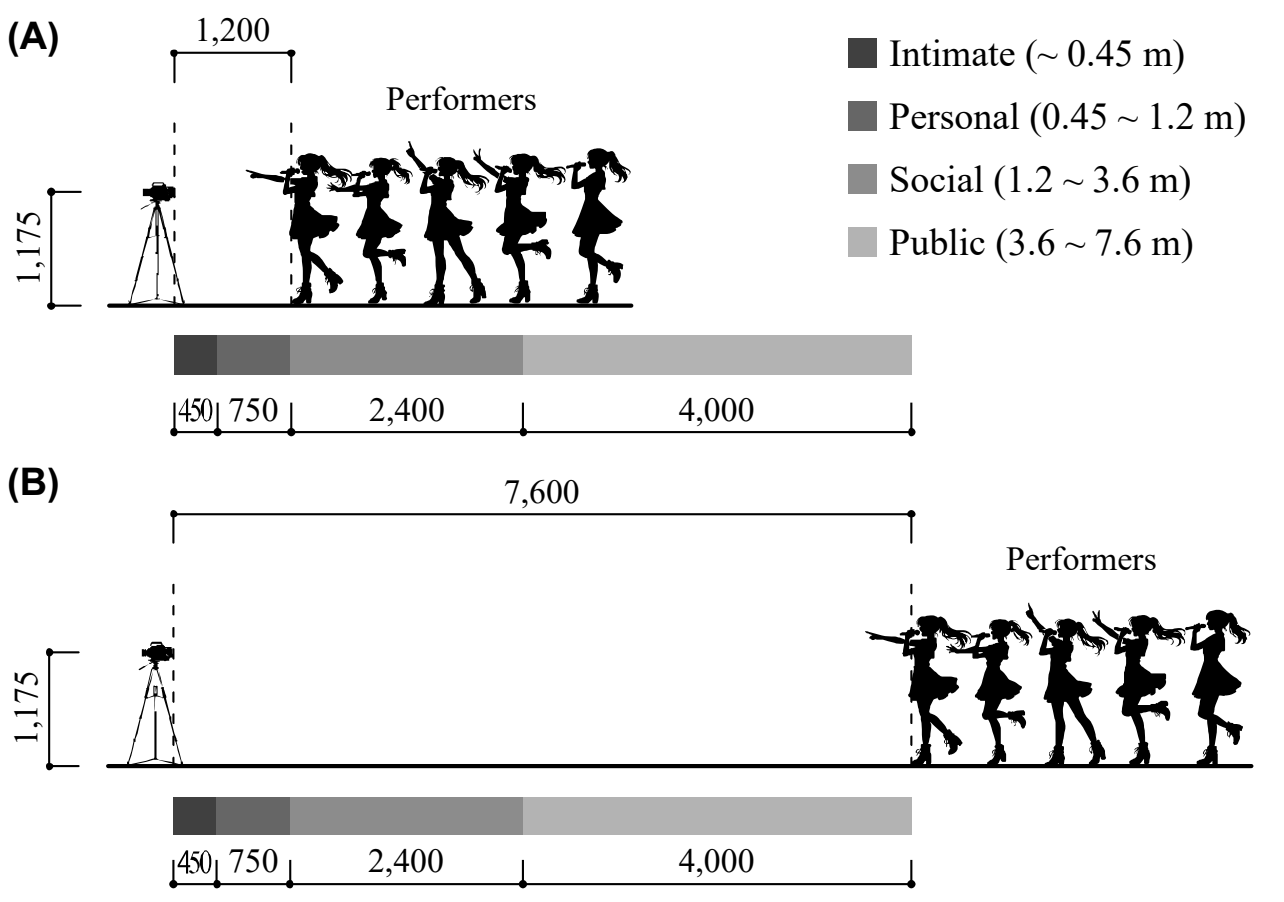

**FIGURE 3** Filming-distance manipulation based on proxemic zones. **(A)** High-Presence condition (1,200 mm). **(B)** Low-Presence condition (7,600 mm).

Accordingly, in the present study, Presence related to Place Illusion in immersive video was manipulated through filming distance, and we examined how this experimental difference altered recipients' fan intensity towards the content target. Specifically, 180° stereoscopic immersive video of an idol live performance was used, and high-Presence and low-Presence conditions were established by manipulating the filming distance from the performers (Fig. 3). This design made it possible to compare differences in the strength of Place Illusion arising from filming distance while holding constant the performers, song, and presentation format.

Furthermore, the present study also focused on the possibility that the degree of prior involvement with the target content might modulate the psychological consequences of immersive video experience. That is, even when the same immersive video is experienced, changes in fan intensity may differ between recipients who already have a high level of involvement with the target content, namely Avid fans, and those whose involvement remains relatively low, namely Casual fans. In particular, whereas Avid fans may already have high fan intensity before the experience, Casual fans may have relatively greater room for change through enhanced Presence. To examine this possibility, the present study included participant groups with different levels of prior involvement and compared changes in fan intensity before and after the experience.

The distinctive feature of the present study is that it moves beyond a descriptive evaluation of whether immersive video is "highly immersive" and explicitly asks, on the basis of Presence theory, which aspects of Presence are associated with what kind of change in fan intensity. In other words, this study attempts to connect Presence research grounded in perceptual and experimental psychology with the formation of relations with content targets, and particularly with the deepening of fan intensity. Through this approach, we empirically examined how experience design that enhances Presence may increase the extent to which a target is experienced as close to the self.

# 2 METHODS

## 2.1 Production of the Immersive Video

### 2.1.1 Content

In the present study, a live performance by idols was adopted as content through which changes in fan intensity before and after the immersive video experience could be measured. STU48[*1], a Japanese idol group, was selected as the target content because it was possible to recruit a sufficient number of participants with different levels of prior involvement. The stimulus video was a newly filmed live performance of the group's song "Shukkō", performed by 12 members, with a duration of approximately 3 min 17 s.

The high-Presence and low-Presence conditions used the same song, performers, choreography, and venue. The performance itself was highly reproducible, as both the choreography and vocal-line assignments had been predetermined. The principal factors that differed between conditions were therefore the position of the filming camera and the position of the audio-recording microphone. The interlude included an unspecified choreographic section in which the performers were free to appeal to the audience seats. This section was also directed in advance so that the staging would be comparable across the two conditions.

### 2.1.2 Presence Manipulation

In the present study, filming distance was set as the experimental variable for manipulating Presence related to Place Illusion. On the basis of proxemic zone categories (Hall, 1966) and prior work on camera distance in cinematic virtual reality (Probst, Rothe, & Hussmann, 2021), the near-distance condition was defined as the high-Presence condition, whereas the distant-view condition was defined as the low-Presence condition.

Filming distance was defined with reference to the standing position of the performer at the centre position. In the high-Presence condition, the camera was placed 1,200 mm from this reference point, whereas in the low-Presence condition it was placed 7,600 mm away (Fig. 4). According to the proxemic zone definitions, the front-row performers in the high-Presence condition were positioned at the distal end of Personal distance, whereas those in the low-Presence condition were positioned at the distal end of Public distance. In addition, when the 12 performers stood in a horizontal line at the front of the stage, the maximum horizontal field of view was approximately 133.6° in the high-Presence condition and approximately 40.5° in the low-Presence condition. Further details of the filming distance are provided in the Supplementary Materials.

*1 STU48 is a Japanese large-scale idol group formed in 2017 and is one of the regional groups within the female idol group network centred on AKB48. As an approximate indication of the size of its fan community, the group's official accounts had 105k followers on X, 96k subscribers on YouTube, and 139k followers on Instagram as of March 2026. The 12 performers who participated in the filming were fourth-generation trainees who had joined the group approximately six months earlier.

The filming height was set at 1,175 mm so that, when the immersive video was experienced in a seated posture, no substantial mismatch would arise in the relative height between the viewer and the performers. This value was determined with reference to anthropometric dimensions for seated Japanese individuals (Ministry of Land, Infrastructure, Transport and Tourism, 2025).

#### 2.1.3 Shooting Conditions

The filming was conducted without an audience in a live venue with a capacity of approximately 300 people.*2 After filming at the high-Presence position, the equipment was moved and filming was then conducted at the low-Presence position.

A Blackmagic URSA Cine Immersive camera (Blackmagic Design) was used, equipped with a Blackmagic 210° Fisheye lens and an inter-lens distance of 64 mm between the centres of the left and right lenses. Detailed camera settings during filming are provided in the Supplementary Materials.

Audio was recorded using a first-order Ambisonics B-format microphone (Zoom H3-VR; Zoom) mounted on top of the camera. The musical performance and vocals were presented using an electroacoustic system: the members' live vocals were mixed on site with the instrumental backing track and played back through the loudspeakers installed in the venue.*3 With respect to sound pressure level, the gain was adjusted while preserving the relative differences in sound pressure between the filming positions.

#### 2.1.4 Editing and Post-Production

The filmed footage was edited using DaVinci Resolve 20.3.2 (Blackmagic Design). To avoid staging-related differences that might affect Plausibility Illusion, no cuts were introduced, and the entire song was edited as a single take. In addition, 5 s of black screen was inserted before and after the video to provide sufficient operational margin during the experiment.

In the high-Presence condition, the filming scaffold was visible when looking downwards; therefore, cropping was applied only to the corresponding area. The peripheral edge blend was also trimmed to the extent required to prevent the left and right lens areas from being visible. Other minor editing procedures are described in the Supplementary Materials.

The export settings were 4,320 × 4,320 px per eye, a frame rate of 90 fps, a bit rate of 150 Mbps, and ASAF as the audio format. The file was exported in the .aivu format.

#### 2.1.5 Implementation of the Stimulus Applications

The exported .aivu files were implemented for presentation on Apple Vision Pro (M2; Apple) using AVPlayer (AVFoundation / AVKit), which supports immersive video playback. The audio was presented as Ambisonics-based three-degree-of-freedom spatial audio. The high-Presence and low-Presence videos were each built as separate applications for Apple Vision Pro.

### 2.2 Participants

Participants were recruited through the official social media accounts of STU48. Recruitment announcements were posted via X and Instagram Stories, and applicants who met the participation criteria were selected. The pre-screening criteria were as follows: participants had to be native speakers of Japanese, have unaided or contact-lens-corrected normal vision (decimal visual acuity ≥ 1.0), have no visual or auditory impairments, not be taking medication that could induce drowsiness, and show low tendencies towards social anxiety.

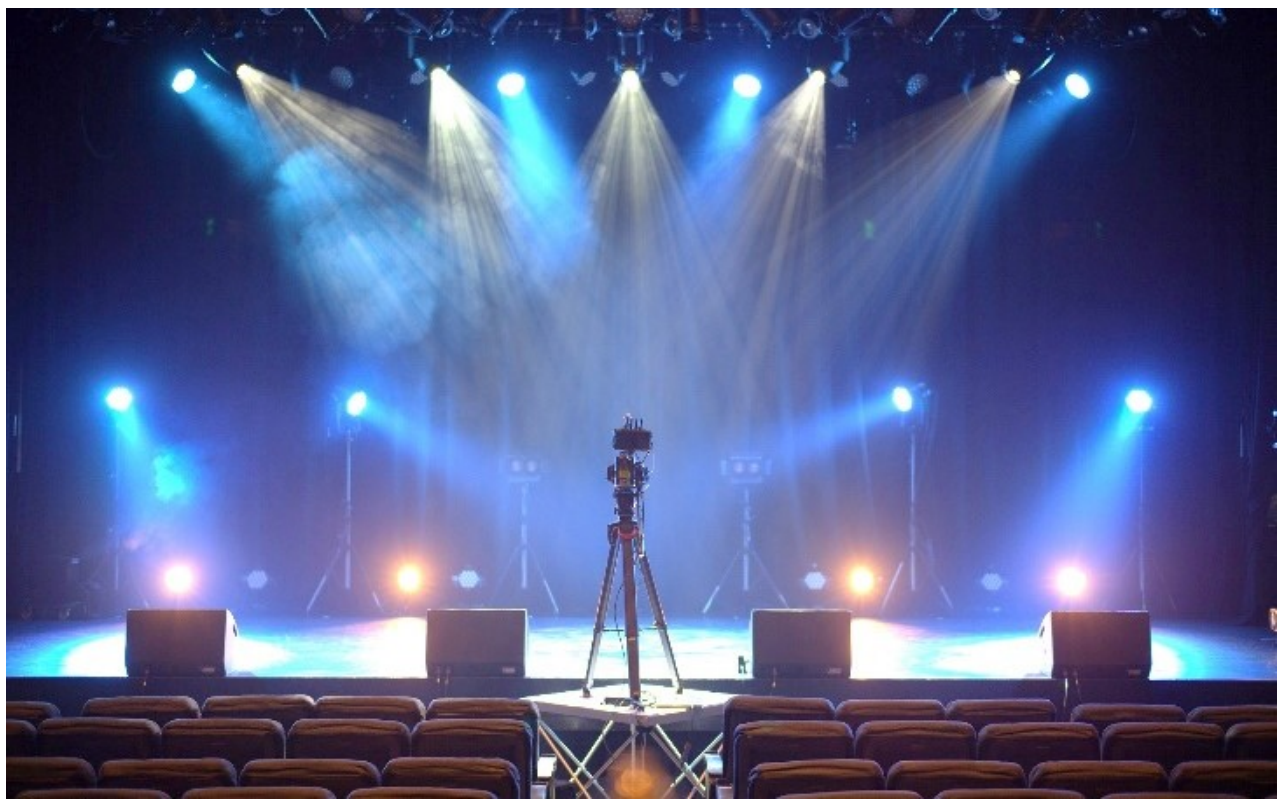

**FIGURE 4** Photograph taken from the low-Presence viewing distance, showing the camera placement for the high-Presence condition.

Given that the high-Presence condition involved video in which participants could feel as though they were making eye contact with the performers at close range, tendencies towards social anxiety were checked from the perspectives of social anxiety (Nagata *et al.*, 2013), tolerance of gaze using a shortened and modified version of the Gaze Anxiety Rating Scale (Langer *et al.*, 2014), and interpersonal distance (Iachini *et al.*, 2014).

The application questionnaire included the item, "Are you a member of the STU48 fan club?" In the present study, fan-club membership status was used as an operational indicator of the degree of prior involvement with the target content. Specifically, fan-club members were defined as Avid fans, representing the high-involvement group, whereas non-members were defined as Casual fans, representing the low-involvement group.

Twenty-five participants took part in the experiment. One participant was excluded because they were not a native speaker of Japanese; therefore, the final analytical sample comprised 24 participants. This sample consisted of 12 Casual fans (6 male, 6 female; age: 35.4 ± 13.1 years, M ± SD) and 12 Avid fans (12 male, 0 female; age: 46.8 ± 12.8 years, M ± SD).

### 2.3 Experimental Procedure

The experiment consisted of five stages, with approximate durations as follows: a Baseline questionnaire (5 min), the first immersive video experience (4 min), a Post1 questionnaire (5 min), the second immersive video experience (4 min), and a Post2 questionnaire (20 min).

The experiment was conducted using three experimental booths in which a quiet environment was ensured. The start times were staggered so that participants would not come into contact with one another during the experiment or immediately before or after it. After entering the experimental booth, participants received an explanation from the experimenter and provided informed consent. They then completed the Baseline questionnaire, put on Apple Vision Pro, and experienced either the high-Presence or low-

*2 Shibuya Pleasure Pleasure, Tokyo; 318 seats.

*3 Although an electroacoustic system was used, in the high-Presence condition, live acoustic sounds were also recorded, such as the performers' footsteps as they changed dance formations.

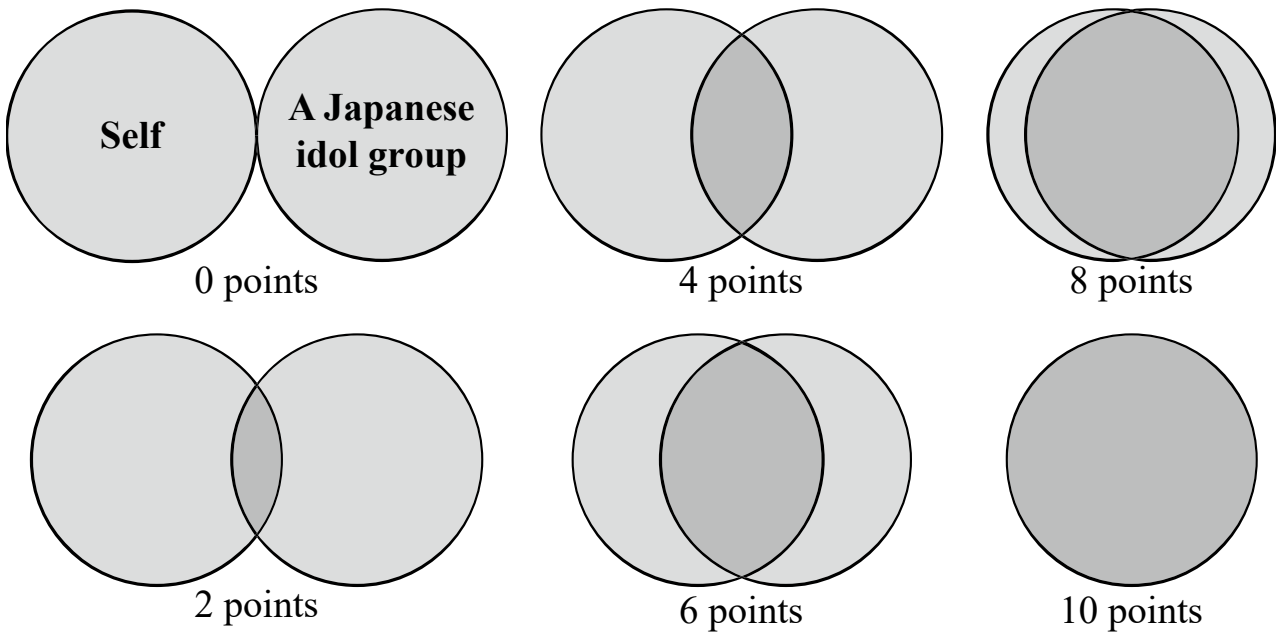

**FIGURE 5** 11-point IOS scale (note: in practice, unlabeled, unshaded circles were used).

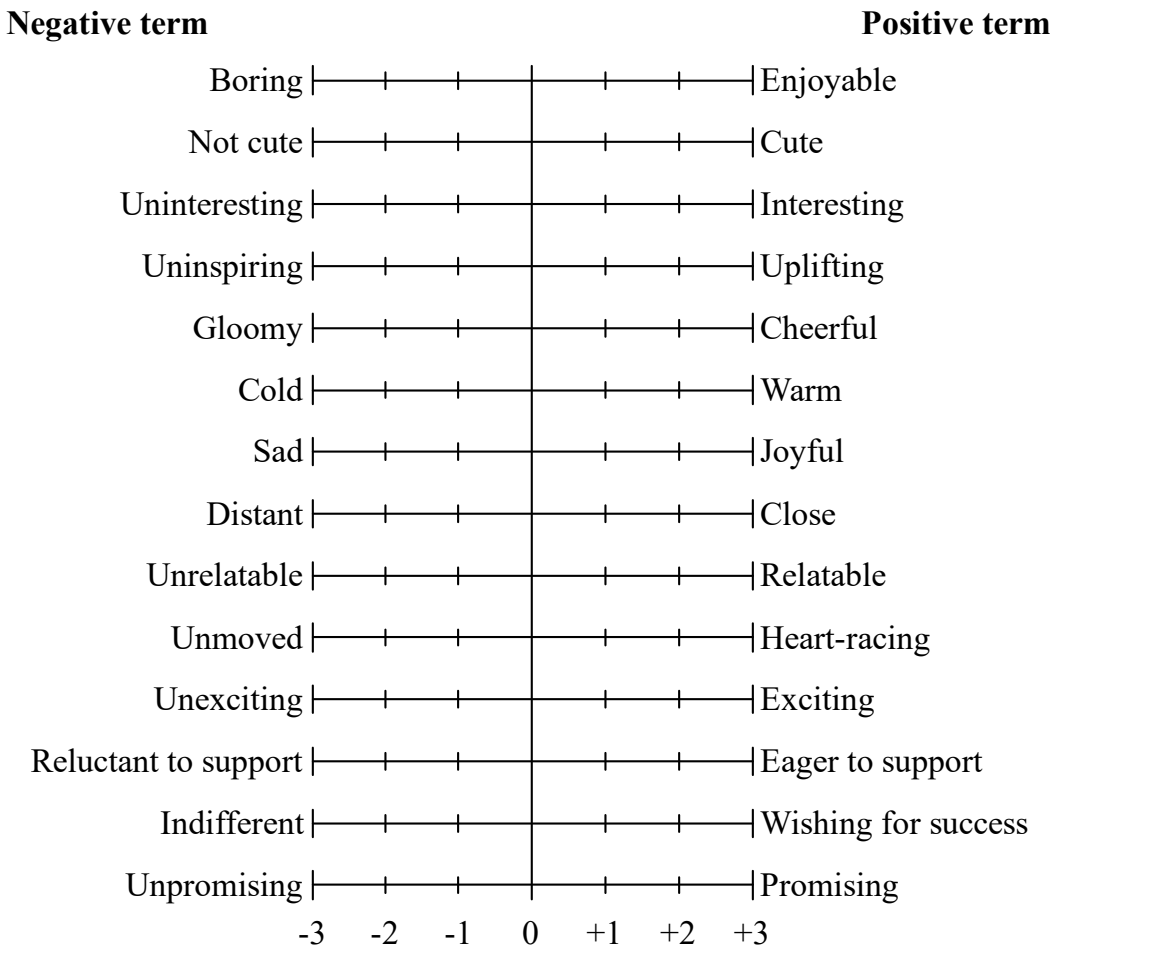

**FIGURE 6** Fourteen adjective pairs used in the semantic differential (SD) method, arranged with the negative term on the left and the positive term on the right. The term "Cute" corresponds to the Japanese adjective *kawaii* in the original questionnaire.

Presence video as Condition 1. After the experience, they removed Apple Vision Pro and completed the Post1 questionnaire using an iPad. Subsequently, they put on Apple Vision Pro again and experienced the other video, which they had not experienced in Condition 1, as Condition 2. Finally, they removed Apple Vision Pro and completed the Post2 questionnaire.

The presentation order of the videos was counterbalanced within each prior-involvement group, namely the Avid fan group and the Casual fan group. Specifically, within each group, equal numbers of participants were assigned to experience the high-Presence video first and the low-Presence video first.

The experiment was conducted after ethical review as a human-subject experiment, in accordance with the third edition of the ethical principles of the Japanese Psychological Association (Japanese Psychological Association, 2011), and after internal approval had been obtained. Participants were also informed in advance that the experiment would be terminated immediately if symptoms such as cybersickness were observed.

## 2.4 Design of Questionnaire Items

In the present study, responses were collected at three time points: the Baseline questionnaire, the Post1 questionnaire, and the Post2 questionnaire. At Baseline, Post1, and Post2, the IOS scale, which served as the primary measure, and Current Affective State, which served as an exploratory measure, were repeatedly administered in order to examine changes before and after the experience. At Post1 and Post2, Core Presence and the subcomponents of Presence were measured with reference to the immersive video that participants had just experienced. In addition, at Post2, items related to experience quality, adverse effects and potential confounds, participant characteristics, technology experience, affinity for technology, and auxiliary items relevant to content business research were collected. A full list of the questionnaire items used in the experiment is provided in the Supplementary Materials.

### 2.4.1 Fan Intensity

In the present study, fan intensity was operationally defined as psychological closeness to the content target and was measured using the Inclusion of Other in the Self (IOS) scale (Aron *et al.*, 1992). The IOS scale is a single-item measure that visually and intuitively assesses psychological closeness between the self and another by means of the degree of overlap between two circles. In this study, an 11-point extended version proposed as IOS11 was used as a reference (Baader *et al.*, 2024). The circle on the left represented the participant, and the circle on the right represented STU48. Participants were instructed as follows: "Please indicate, using the overlap between the circles, how psychologically close you currently feel yourself to be to STU48." They then rated their current subjective fan intensity on an 11-point scale ranging from 0 to 10 (Fig. 5). In the actual response screen, circle diagrams without labels or shading were used.

### 2.4.2 Current Affective State

To capture momentary affective states towards the content target, impression ratings were administered as the Current Affective State measure using the semantic differential (SD) method (Osgood *et al.*, 1957). The adjective pairs were developed through a preliminary free-response survey administered to a panel of respondents who supported a specific idol target, not limited to STU48 ($n = 49$). Respondents were asked to provide adjectives that they felt were "currently appropriate" and "currently inappropriate" for describing the idol target they supported. These responses were preliminarily organised by the authors, after which 14 conceptually paired items were selected.

For each adjective pair, the negative term was placed on the left and the positive term on the right, and responses were collected on a seven-point scale ranging from −3 to +3 (Fig. 6). This measure was used to exploratorily examine the structure of liking, perceived closeness, emotional arousal, intention to support, and related affective responses towards the content target.

### 2.4.3 Presence Measures

Presence was assessed at Post1 and Post2 with reference to the immersive video that participants had just experienced. To confirm whether the filming-distance manipulation had produced differences in Presence related to Place Illusion, Spatial Presence, the Slater–Usoh–Steed questionnaire (SUS), and Self-Location were measured as Core Presence indices (Slater *et al.*, 1994; Schubert *et al.*, 2001; Slater, 2009).

Spatial Presence was assessed with a single item asking the extent to which participants felt that they had been there. The SUS-related items assessed the extent to which the video world became the dominant reality and whether the experience felt closer to visiting a place than merely watching a video. Self-Location was assessed with two items asking whether participants felt that they were in the midst of the events and part of the video experience.

In addition, the following subcomponents of Presence were measured: Social Presence, for example, “Did you feel that you wanted to make eye contact with the performers?”; Possible Actions, for example, “Did you feel that you could touch the performers if you reached out your hand?”; Perceptual Realism, for example, “Do you think the atmosphere and smell in the video would feel the same as if they were experienced at an actual live venue?”; Engagement, for example, “Did you ever feel as though you were about to vocalise in response to the performers?”; and Observability, for example, “To what extent could you observe the performers’ facial expressions?” Social Presence was defined as a two-item composite index capturing the sense of the presence of others in a mediated environment (Biocca *et al.*, 2003; Lombard & Ditton, 1997). Possible Actions was defined as a two-item composite index concerning the possibility of action within the environment, corresponding to a subdimension of Spatial Presence (Wirth *et al.*, 2007; Hartmann *et al.*, 2016). Perceptual Realism was defined as a two-item composite index corresponding to Ecological Validity in presence scales such as the ITC-Sense of Presence Inventory (Lessiter *et al.,* 2001). Engagement was defined as a three-item composite index capturing vocal response, smiling response, and absorption in the experience. Observability was not a standard factor in existing scales, but was defined as an original two-item auxiliary index in the present study to capture visual access to the performers associated with the filming-distance manipulation. The full questionnaire items are provided in the Supplementary Materials.

### 2.4.4 Auxiliary Items and Potential Confounds

At Post2, after all video experiences had been completed, participants provided overall evaluations concerning behavioural intentions, experience quality, adverse effects and potential confounds, comparative and content evaluations, demographic characteristics, technology experience and affinity for technology, and willingness to participate in follow-up surveys. Specifically, willingness to pay (WTP), Net Promoter Score (NPS), intention to attend a live performance in person, intention to post user-generated content (UGC), satisfaction, enjoyment, and intention to re-experience similar content were collected.

In addition, with reference to the Simulator Sickness Questionnaire, factors that could interfere with concentration on the video experience were assessed, including cybersickness, eye fatigue, headache, discomfort, the fit and feel of Apple Vision Pro, image quality, sound quality, and the experimental environment (Kennedy *et al.*, 1993). Demographic and background items included gender, age, characteristics of fan activities, experience with XR devices, experience with Apple Vision Pro, and prior experience of immersive video. The nine-item Affinity for Technology Interaction scale was also administered (Franke *et al.*, 2019).

In the present paper, items related to content business research, such as state-dependent intentions, WTP, NPS, intention to attend a live performance in person, and UGC intention, were not included in the main analyses. This decision was made because these items depend substantially on the specific context of the experienced content and should therefore be analytically distinguished from the main focus of this paper, namely the relationship between Presence and fan intensity.

## 2.5 Analysis

Post1 and Post2 responses were reorganised according to each participant’s experience order and assigned to the video condition experienced immediately before each questionnaire, yielding data for the high-Presence and low-Presence conditions. For the IOS scale, raw scores were used in analyses examining the absolute level of fan intensity, whereas difference scores calculated as Post − Baseline ($\Delta$IOS scores) were used in analyses examining changes induced by the experience.

Before all parametric tests, the distributions of the main dependent variables and model residuals were examined using Shapiro–Wilk tests and Q–Q plots (Shapiro & Wilk, 1965). In analyses involving repeated-measures factors with three or more levels, sphericity was assessed using Mauchly’s test, and Greenhouse–Geisser correction was applied where necessary (Mauchly, 1940; Greenhouse & Geisser, 1959). For multiple comparisons, p values were adjusted using Holm’s method (Holm, 1979). On the basis of the distributional checks, IOS scale scores and seven-point rating items were treated as approximately continuous variables. The significance level was set at $\alpha = .05$, and $\eta p^2$ was reported as the effect size for analyses of variance.

The effect of experience order was examined using a linear mixed model (LMM) with $\Delta$IOS score as the dependent variable (Laird & Ware, 1982). The fixed effects were prior-involvement group, Presence condition, experience order, and the interaction between prior-involvement group and Presence condition. A random intercept was specified for each participant. The conceptual model formula was as follows:

*ΔIOS score ~ Prior involvement group + Presence condition + Experience order + Prior involvement group × Presence condition + (1 | Participant)*

For the Presence measures, the high-Presence and low-Presence conditions were compared as a manipulation check for the filming-distance manipulation. Paired-samples *t* tests were conducted for the Core Presence indices, namely Spatial Presence, SUS, and Self-Location. Condition comparisons were also conducted in the same manner for the subcomponents of Presence, namely Social Presence, Possible Actions, Perceptual Realism, Engagement, and Observability. In addition, to examine whether Presence ratings depended on prior-involvement group, a supplementary mixed two-way analysis of variance (ANOVA) was conducted, with Presence condition as a within-participants factor and prior-involvement group, namely the Avid fan group and the Casual fan group, as a between-participants factor.

For the IOS scale, two types of analysis were conducted: raw-score analysis and $\Delta$IOS-score analysis. First, to examine the absolute level of fan intensity, a 2 × 3 mixed two-way ANOVA was conducted with raw score as the dependent variable, prior-involvement group (Avid fans / Casual fans) as a between-participants factor, and evaluation timing (Baseline / Post low-Presence / Post high-Presence) as a within-participants factor. Second, to examine changes induced by the experience, a 2 × 2 mixed two-way ANOVA was conducted with $\Delta$IOS score as the dependent variable, prior-involvement group as a between-participants factor, and Presence condition (low-Presence / high-

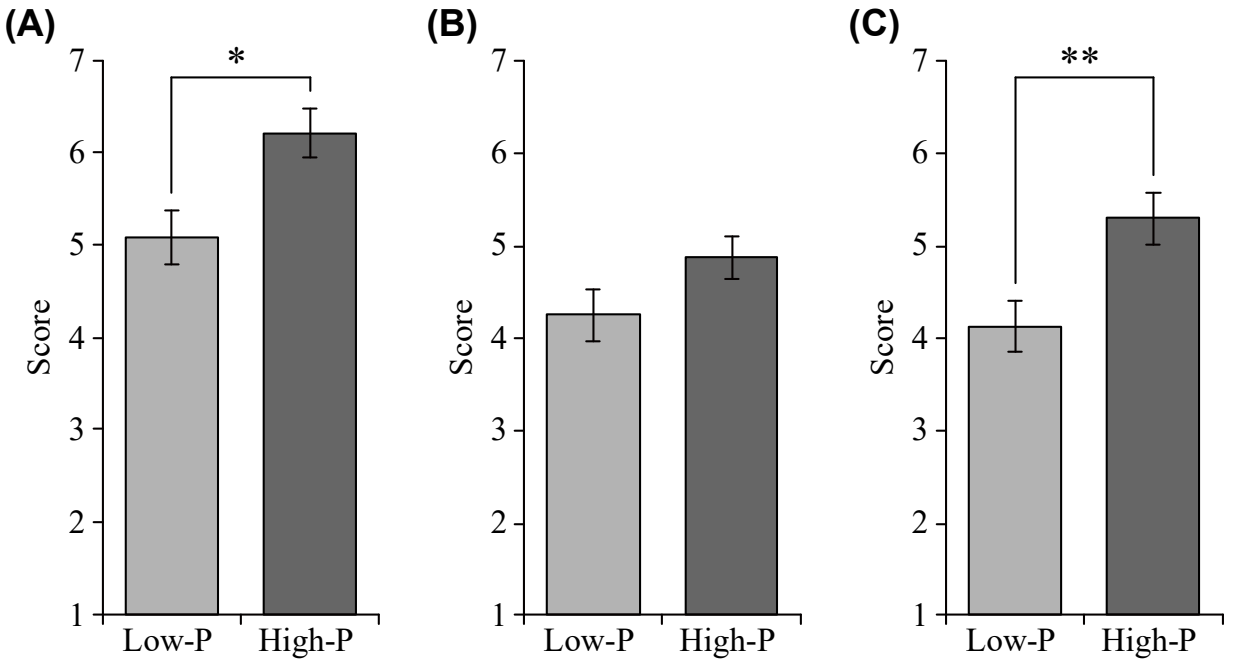

**FIGURE 7** Manipulation check of core Presence measures. Mean scores for **(A)** Spatial Presence, **(B)** Slater–Usoh–Steed questionnaire (SUS), and **(C)** Self-Location in the low-Presence (Low-P) and high-Presence (High-P) conditions. Error bars indicate standard errors.

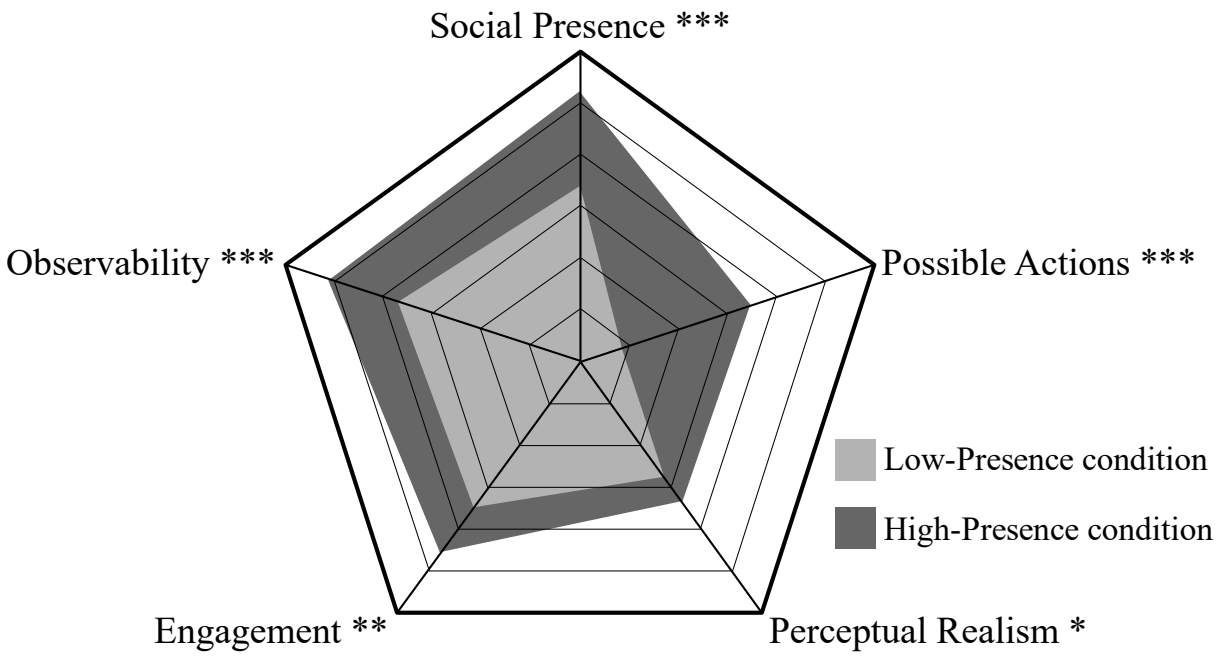

**FIGURE 8** Subdimensions of Presence in the low- and high-Presence conditions. Mean scores for Social Presence, Possible Actions, Perceptual Realism, Engagement, and Observability are shown for each condition. Higher values indicate stronger perceived Presence-related experience.

Presence) as a within-participants factor. When the interaction was not significant, simple main-effect tests by prior-involvement group were not treated as part of the main analysis.

In addition, supplementary analyses were conducted to examine whether the IOS scale scores of Casual fans after the high-Presence video experience reached a level comparable to the Baseline IOS scale scores of Avid fans. The Post high-Presence scores of the Casual fan group and the Baseline scores of the Avid fan group were treated as two independent groups, and Welch's two-sample *t* test was conducted. Furthermore, because the absence of a significant difference does not constitute evidence of equivalence, an equivalence test using the two one-sided tests (TOST) procedure was also conducted (Schuirmann, 1987; Lakens, 2017). The equivalence margin was set at a standardised effect size of $d = \pm 0.5$.

For Current Affective State, exploratory factor analyses were conducted separately at Baseline, Post low-Presence, and Post high-Presence using the 14 SD-method items. Previous studies have used procedures in which multi-timepoint SD-method data are factor-analysed separately at each time point and attitude or impression structures are compared through differences in factor-loading patterns (Terashita *et al.*, 2016). Changes in factor structure before and after an intervention have also been discussed within the framework of *response shift*, particularly *gamma change* (Golembiewski *et al.*, 1976; Jabrayilov *et al.*, 2017). On this basis, the present study conducted timepoint-specific factor analyses as supplementary analyses to exploratorily examine structural changes in Current Affective State. The number of factors was determined by considering scree plots, the eigenvalue-greater-than-one criterion, and interpretability (Fabrigar *et al.*, 1999; Costello & Osborne, 2005). After factor extraction, oblique rotation was applied, and items with factor loadings of 0.4 or above were treated as the main targets of interpretation. In addition, hierarchical clustering and scatter plots in the factor space were created to supplementarily examine relations among items.

All analyses were conducted in Python.

# 3 RESULTS

## 3.1 Analysis of Experience-Order Effects

Because the present experiment employed a within-participants design in which participants experienced both the high-Presence and low-Presence videos, the effect of experience order was examined using a linear mixed model before the main analyses were conducted. The dependent variable was $\Delta$IOS score, and prior involvement, Presence condition, experience order, and the interaction between prior involvement and Presence condition were entered as fixed effects.

The results showed that only the fixed effect of Presence condition was significant ($\beta$ = 3.083, SE = 0.712, $z$ = 4.331, $p < .001$). By contrast, the effect of experience order was not significant ($\beta$ = 0.917, SE = 0.663, $z$ = 1.382, $p$ = .167). The effect of prior involvement ($\beta$ = 1.167, SE = 0.833, $z$ = 1.401, $p$ = .161) and the interaction between prior involvement and Presence condition ($\beta$ = −1.167, SE = 1.007, $z$ = −1.159, $p$ = .247) were also not significant.

Accordingly, in the subsequent analyses, no effect of experience order was assumed, and the high-Presence and low-Presence conditions were treated as the condition factor.

## 3.2 Manipulation Check of Presence

To confirm whether the filming-distance manipulation produced differences in subjective Presence, Presence ratings in the high-Presence and low-Presence conditions were compared using paired-samples t tests (Fig. 7). Among the Core Presence indices, Spatial Presence and Self-Location were significantly higher in the high-Presence condition than in the low-Presence condition (Spatial Presence: $t_{(23)} = 2.873, p < .05$; Self-Location: $t_{(23)} = 3.536, p < .01$). By contrast, SUS did not differ significantly between conditions ($t_{(23)} = 1.983, p = .059$).

The high-Presence condition also consistently exceeded the low-Presence condition for the subcomponents of Presence. Significant differences were observed for Social Presence ($t_{(23)} = 6.511$, $p < .001$), Possible Actions ($t_{(23)} = 11.720$, $p < .001$), Perceptual Realism ($t_{(23)} = 2.837, p < .05$), Engagement ($t_{(23)} = 3.649, p < .01$), and Observability ($t_{(23)} = 6.000$, $p < .001$). In particular, the differences were large for Possible Actions, Social Presence, and Observability, indicating that the high-Presence condition strongly evoked a sense of possible action within the recorded space, social presence, and visual access to the performers (Fig. 8).

Neither the main effect of prior involvement on Presence ratings nor the interaction between prior involvement and Presence condition was significant. Thus, the filming-distance manipulation of Presence was established similarly for both the Avid fan and Casual fan groups. Compared with the low-Presence condition, the high-Presence condition in the present study functioned as a manipulation that broadly enhanced Presence-related experience, with theoretically central increases in Self-Location and particularly large condition differences in Possible Actions, Social Presence, and Observability.

## 3.3 Effects of Presence on IOS scale

### 3.3.1 Raw Score Analysis

Using the raw IOS scale scores, differences in the absolute level of fan intensity were examined as a function of prior involvement and evaluation timing (Table 1). A 2 × 3 mixed two-way ANOVA was conducted with prior-involvement group (Avid fans / Casual fans) as a between-participants factor and evaluation timing (Baseline / Post low-Presence / Post high-Presence) as a within-participants factor. The main effect of prior involvement was significant ($F_{(1, 22)} = 13.40$, $p = .0014$, $\eta p^2 = .379$). Thus, the Avid fan group showed higher IOS scale scores overall than the Casual fan group (Fig. 9a).

The main effect of evaluation timing was also significant ($F_{(2, 44)} = 32.01$, $p < .001$, $\eta p^2 = .593$). By contrast, the interaction between prior involvement and evaluation timing was not significant ($F_{(2, 44)} = 1.12$, $p = .336$, $\eta p^2 = .048$).

These results confirmed that the absolute level of fan intensity differed according to prior-involvement group. However, because raw scores included group differences that were already present before the experience, increases induced by the experience were examined using ⊿IOS scores.

### 3.3.2 ⊿IOS Score Analysis (Difference Scores)

To examine increases in fan intensity induced by the experience, a 2 × 2 mixed two-way ANOVA was conducted with ⊿IOS score as the dependent variable, prior involvement as a between-participants factor, and Presence condition as a within-participants factor (Table 2). The main effect of Presence condition was significant ($F_{(1, 22)} = 24.66$, $p < .001$, $\eta p^2 = .529$), indicating that ⊿IOS scores after the high-Presence video experience were significantly higher than those after the low-Presence video experience (Fig. 9b).

By contrast, neither the main effect of prior involvement ($F_{(1, 22)} = 0.743$, $p = .398$, $\eta p^2 = .033$) nor the interaction between prior-involvement group and Presence condition ($F_{(1, 22)} = 1.34$, $p = .259$, $\eta p^2 = .058$) was significant.

Taken together, these results indicate that although the absolute level of fan intensity differed by prior-involvement group, the increase in fan intensity induced by the experience corresponded to Presence condition. In other words, compared with the low-Presence video, the high-Presence video produced a greater increase in fan intensity towards the content target.

## 3.4 Comparison Between Casual Fans After the High-Presence Experience and Avid Fans Before the Experience

A supplementary analysis was conducted to examine whether the IOS scale scores of the Casual fan group after the high-Presence video experience approached the pre-experience level of the Avid fan group. The Baseline score of the Avid fan group was 5.25 ± 2.26 (M ± SD), whereas the Post high-Presence score of the Casual fan group was 5.50 ± 2.71. Welch's two-sample *t* test showed that the difference between the two groups was not significant ($t_{(21.31)} = -0.245$, $p = .809$).

However, TOST with an equivalence margin of $d = \pm 0.5$ did not support equivalence. The lower-bound test was $t_{(21.31)} = 0.980$, $p = .169$, and the upper-bound test was $t_{(21.31)} = -1.470$, $p = .078$. Therefore, although the IOS scale scores of Casual fans after the high-Presence video experience were not significantly lower than the pre-experience scores of Avid fans, the two scores cannot be regarded as statistically equivalent.

**TABLE 1** Raw IOS scores were analysed using a 2 × 3 mixed ANOVA with prior involvement group as a between-participants factor and evaluation timing (Baseline, Post low-Presence condition, Post high-Presence condition) as a within-participants factor.

| Source | *df* | *F* | *p* | $\eta p^2$ |
|---|---|---|---|---|
| Prior involvement group | 1, 22 | 13.40 | .0014 | .379 |
| Evaluation timing | 2, 44 | 32.01 | < .001 | .593 |
| Prior involvement group × Evaluation timing | 2, 44 | 1.12 | .336 | .048 |

**TABLE 2** ⊿IOS scores (Post − Baseline) were analysed using a 2 × 2 mixed ANOVA with prior involvement group as a between-participants factor and Presence condition (low-Presence vs. high-Presence) as a within-participants factor.

| Source | *df* | *F* | *p* | $\eta p^2$ |
|---|---|---|---|---|
| Prior involvement group | 1, 22 | 0.743 | .398 | .033 |
| Presence condition | 1, 22 | 24.66 | < .001 | .529 |
| Prior involvement group × Presence condition | 1, 22 | 1.34 | .259 | .058 |

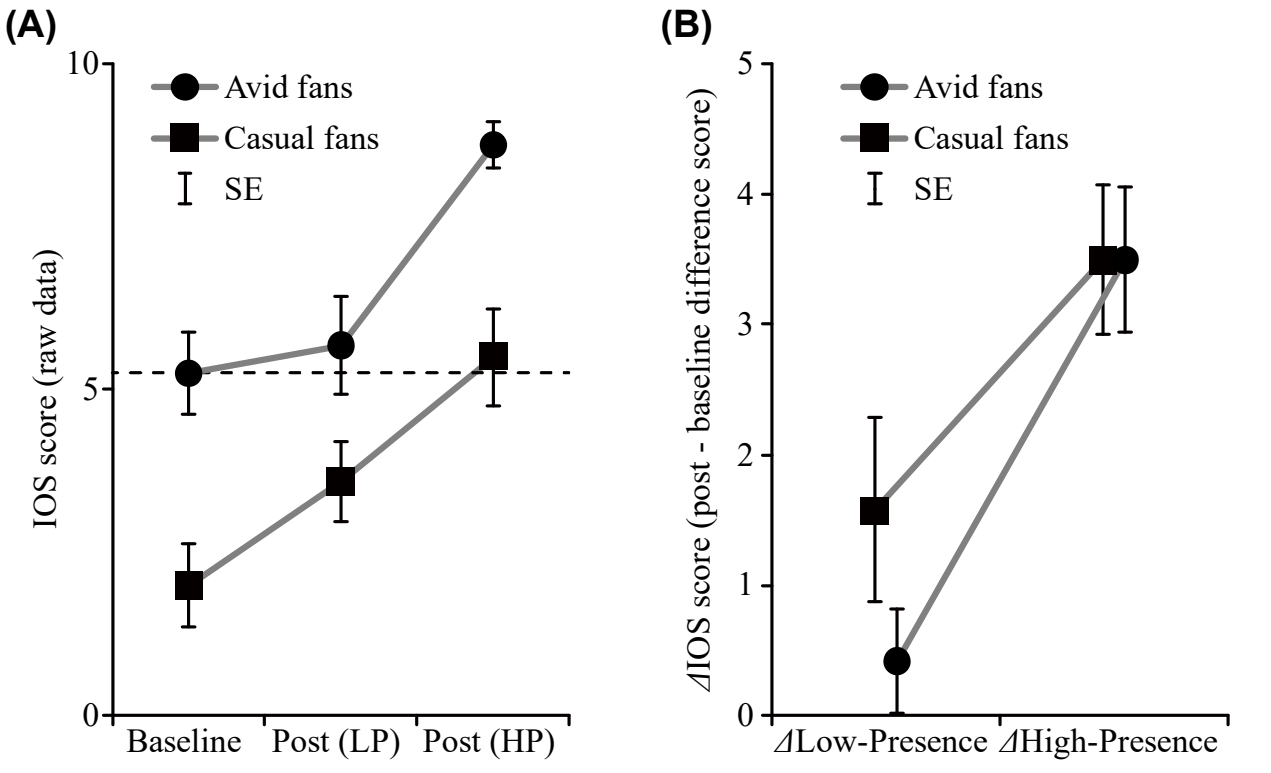

**FIGURE 9** Inclusion of Other in the Self scale results for participants with different levels of fandom **(A)** raw scores at Baseline, Post low-Presence, and Post high-Presence; **(B)** difference scores calculated as Post – Baseline.

## 3.5 Exploratory analysis of Current Affective State

Exploratory factor analyses of Current Affective State were conducted separately at Baseline, Post low-Presence, and Post high-Presence. The number of factors was determined on the basis of scree plots, the eigenvalue-greater-than-one criterion, and interpretability, and oblique rotation was applied. Items with factor loadings of 0.4 or above were treated as the primary targets of interpretation (Fig. 10). The scree plots, factor-loading matrices, and dendrograms are provided in the Supplementary Materials.

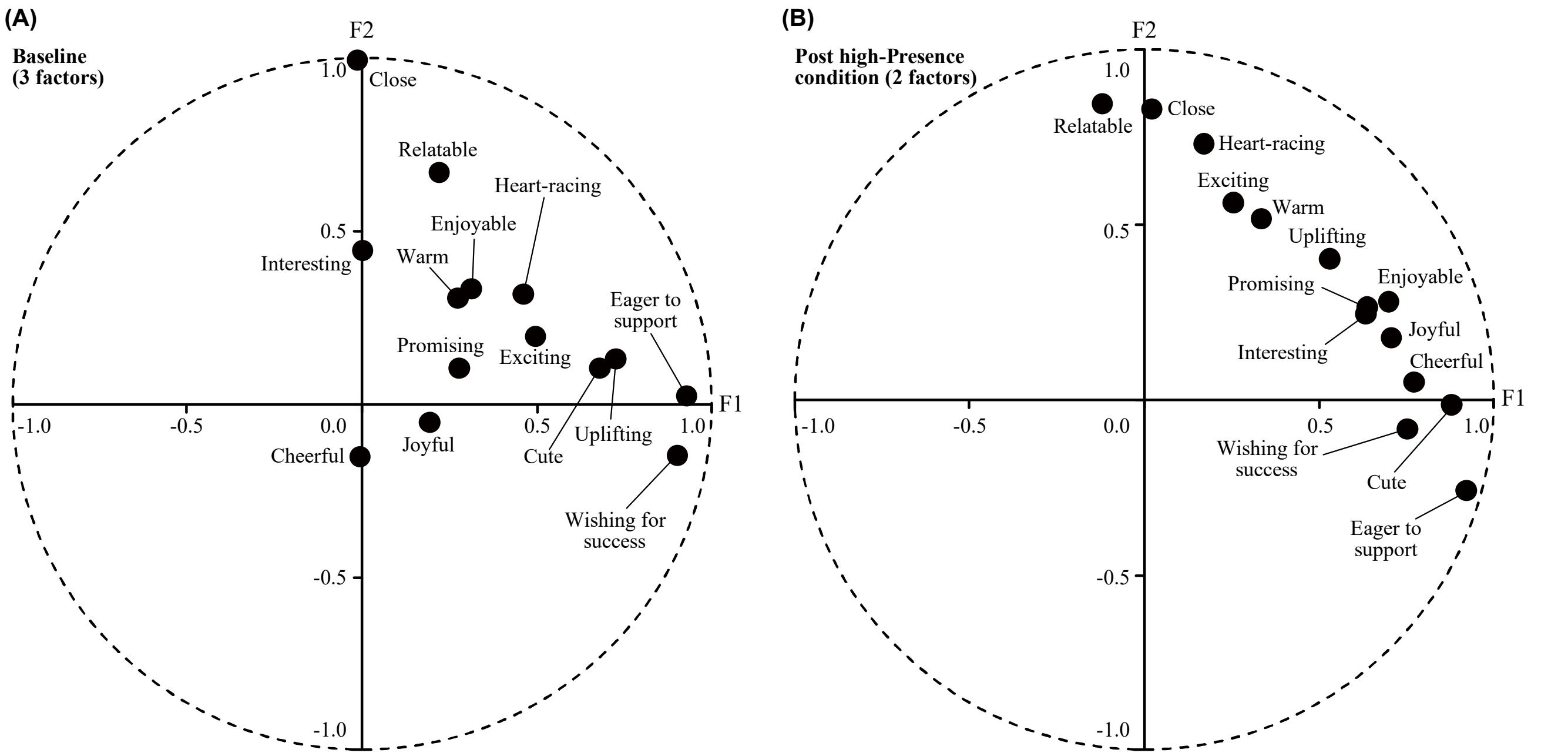

**FIGURE 10** Factor-loading maps of Current Affective State. Panel **(A)** shows the Baseline condition, for which three factors were retained. Panel **(B)** shows the post high-Presence condition, for which two factors were retained. Points indicate affective-state items, and dashed circles indicate unit loading.

At Baseline, a three-factor structure was obtained. Factor 1 was labelled the supportive behavioural tendency factor, with high loadings for "Eager to support", "Wishing for success", "Uplifting", and "Cute". Factor 2 was labelled the psychological closeness factor, with high loadings for "Close" and "Relatable". Factor 3 was labelled the general positive affect factor, with high loadings for "Cheerful", "Joyful", "Interesting", and "Enjoyable" (Fig. 10a). At Post high-Presence, a two-factor structure was obtained. Factor 1 was labelled the overall favourable evaluation factor, because items such as "Eager to support", "Cute", "Cheerful", "Wishing for success", "Joyful", "Enjoyable", "Promising", and "Interesting" showed high loadings. Factor 2 was labelled the closeness and emotional arousal factor, because items such as "Relatable", "Close", "Heart-racing", "Exciting", and "Warm" loaded highly on this factor (Fig. 10b).

At Post low-Presence, the SD items largely converged into a one-factor structure. Because multiple positive evaluation items loaded broadly on this factor, it was labelled the general positive evaluation factor. By contrast, no distinct factor corresponding to the closeness and emotional arousal factor extracted at Post high-Presence was clearly identified at Post low-Presence.

Taken together, the exploratory factor analyses of Current Affective State suggested different factor structures across Baseline, Post low-Presence, and Post high-Presence.

# 4 DISCUSSION

## 4.1 Principal Findings

The high-Presence condition broadly enhanced Presence-related ratings, except for the SUS index, and produced a greater increase in IOS scale Post − Baseline difference scores than the low-Presence condition. Because no experience-order effect was observed, this difference is unlikely to be attributable to presentation order. These results indicate that, at least in the present experiment, the difference between the high-Presence and low-Presence conditions corresponded to differences in Presence produced by filming distance, rather than to the novelty of the head-mounted-display experience itself.

The significance of the present study lies in demonstrating, in live-action immersive video, that enhanced Presence is associated not only with an increased sense of "being there", but also with an increased degree to which the content target is experienced as close to the self. In particular, filming distance functioned not merely as a difference in angle of view or composition, but as a design variable that altered Presence-related experience and corresponded to an increase in fan intensity towards the content target.

These findings provide partial answers to the issues identified in the Introduction. First, the filming-distance manipulation broadly affected Presence-related measures, while the theoretically central increase in Self-Location and the particularly large condition differences in Possible Actions, Social Presence, and Observability indicate that the manipulation was closely tied to the viewer's situated relation to the performers. Second, the increase in fan intensity was not observed as information comprehension or content evaluation, but as a change in fan psychology, namely the extent to which the performers were experienced as close to the self. Thus, the present study suggests that Presence design in immersive video can be connected to the formation of psychological relations with content targets.

## 4.2 Filming distance, Place Illusion, and Self-Location

Although the filming-distance manipulation enhanced most Presence-related ratings, its theoretically important implication concerns Place Illusion. The largest condition differences appeared in aspects closely tied to the viewer's situated relation to the performers, namely Possible Actions, Social Presence, and Observability, alongside a significant increase in Self-Location. The nonsignificant SUS result is not inconsistent with this interpretation, because SUS is a broad Presence measure derived from early research on virtual environments and may be less sensitive to distance manipulations in non-interactive immersive video.

This interpretation suggests the need to treat Presence not as a single comprehensive score, but as a set of subcomponents, including self-location within a space, possible action, and the sense of the presence of others. Spatial Presence is not merely a matter of perceiving the video world as realistic; rather, it concerns the process by which experiencers locate themselves within that space and feel as though some form of action is possible within it (Wirth *et al.*, 2007; Hartmann *et al.*, 2016). Therefore, the increases in Self-Location, Possible Actions, and Social Presence in the high-Presence condition suggest that the filming-distance manipulation did more than simply make the target appear larger. It also formed a spatial and social subjective experience in which participants positioned themselves near the target.

This interpretation is consistent with the theoretical framework that distinguishes Presence into Place Illusion and Plausibility Illusion (Slater, 2009; Slater *et al.*, 2022). In the present study, the performers, song, choreography, venue, presentation device, and video format were controlled across conditions, while filming distance was set as the primary manipulated variable. Accordingly, the condition difference is considered to have arisen primarily from how experiencers perceived their distance from the target in the video, rather than from differences in the semantic coherence of the presented content or in narrative plausibility.

The present findings show that distance design based on Proxemics corresponds not only to Presence ratings in immersive video, but also to psychological closeness to the content target. Whereas previous research on virtual environments has primarily treated comfort, naturalness, Presence, or psychological distance from an avatar as dependent variables, the present study extends this line of work by showing that pseudo-interpersonal distance formed through filming distance may also be related to the extent to which a target is felt to be close to the self. This extension means that interpersonal distance in virtual environments does not merely imitate interpersonal-distance norms in physical space, but may also function as a condition for forming psychological relations with a target. Indeed, even in virtual space, distance from avatars or virtual others has been shown to carry social meaning and to influence interpersonal-distance norms and non-verbal responses (Bailenson *et al.*, 2003; Yee *et al.*, 2007). Similarly, research on camera distance in cinematic virtual reality has reported that distance conditions affect Presence, comfort, naturalness, and psychological distance from avatars (Probst, Rothe, & Hussmann, 2021). The present study extends these findings to psychological closeness to performers in live-action immersive video.

What is important here is that filming distance in immersive video functions differently from shot size or angle of view in ordinary flat video. In flat video, a close shot functions as a compositional technique for making the target appear larger. By contrast, in stereoscopic immersive video presented on a head-mounted display, filming distance specifies where the experiencer's body is felt to be located within the video space. In other words, filming distance is a design variable that simultaneously constitutes not only the amount of visual information, but also pseudo-interpersonal distance from the target, possible action, the perception of gaze, and social closeness.

This interpretation is also consistent with the argument that self-experience in immersive video may be understood as a *self-location-dominant state*, in which self-location at the recorded viewpoint becomes relatively foregrounded (Toida, 2026). In the present study, the increases in Self-Location, Possible Actions, Social Presence, and Observability in the high-Presence condition are consistent with the possibility that filming distance functioned as a design variable that located experiencers near the content target.

## 4.3 Enhancing Presence, Deepening Fan Intensity

The fan intensity measured in the present study was not purchase intention or supportive behaviour itself, but psychological closeness to the content target as assessed using the IOS scale. The IOS scale was developed as a measure that visually assesses psychological overlap between the self and another (Aron *et al.*, 1992; Baader *et al.*, 2024). Therefore, the increase in IOS scale scores observed in the high-Presence condition can be interpreted not merely as indicating that evaluations of the performers became more favourable, but as an index showing that the performers were incorporated as closer entities within the participant's relation to the self.

This result indicates that Presence design in immersive video is not limited to a technical manipulation for presenting the target more vividly. In the high-Presence condition, participants are likely to have located themselves near the content target and to have perceived the performers' facial expressions, gaze, bodily movements, and presence within the space as more proximal. Consequently, the content target may have been experienced not merely as an object of viewing, but as an entity situated closer to the self within a psychological relation.

This interpretation is also consistent with discussions of parasocial interaction and parasocial relationships, in which recipients form one-sided yet intimate relations with people appearing in media (Horton & Wohl, 1956; Rubin *et al.*, 1985; Tukachinsky & Stever, 2019; Schramm *et al.*, 2024). Even in conventional media experiences, gaze, direct address, and repeated exposure can foster familiarity and emotional attachment to a target. However, in immersive video, the target does not appear to exist on the other side of a screen, but rather as being present near the experiencer's body. In this respect, high-Presence immersive video may be positioned as a media form that reorganises "intimacy at a distance" in conventional media into a form of parasocial closeness in proximity, accompanied by spatial self-location. In other words, the increase in fan intensity observed in the present study may not be attributable merely to the effects of gaze or direct address. Rather, it may reflect the spatial and bodily-proximal reconfiguration of a parasocial relation under conditions in which the performers are perceived as if they were near the participant's body.

Taken together, enhanced Presence may function not only to increase the sense of presence in a video experience, but also to reduce psychological distance from the target and deepen fan intensity. In this sense, Presence design in immersive video can be positioned as experience design for forming psychological relations with content targets.

## 4.4 Immersive Intimacy After High-Presence Experience

The exploratory factor analysis using the SD method provided supplementary evidence for the increase in fan intensity observed in the IOS scale, from the perspective of attitudinal and affective structure. After the high-Presence condition, items related to psychological closeness, such as “Close” and “Relatable”, loaded on the same factor as items related to emotional arousal, such as “Heart-racing”, “Exciting”, and “Warm”. This suggests that, in the high-Presence immersive video experience, feeling that the target was “close” may not have been constituted merely as a judgement of distance, but as a relational evaluation accompanied by emotional arousal. In other words, in the high-Presence condition, psychological closeness and affective response may not have appeared as separate evaluative dimensions, but as an integrated subjective state in which the target was emotionally experienced as close to the self.

This state can be interpreted as a form of parasocial psychological closeness in which intimacy towards a person appearing in media emerges in immersive video together with a sense of spatial and bodily-proximal closeness. In immersive video, the target may be perceived not as someone located beyond the screen, but as an entity positioned near the experiencer’s body. This spatial configuration may have made it easier for the sense of the target being close and emotional arousal towards that target to become coupled. In the present study, we provisionally interpret this state as *immersive intimacy*.

In Japanese idol culture, opportunities for contact, such as handshake events, have functioned as practices that increase perceived closeness to idols and the sense of being recognised by them. By contrast, it has been reported that the bodily and affective significance of such face-to-face contact is difficult to reproduce fully in online experiences (Yakura, 2021). Therefore, the present findings should not be taken to mean that immersive video can substitute for bodily contact such as handshake events. Rather, they should be interpreted as suggesting that immersive video may complement part of the psychological function of face-to-face contact, such as perceived closeness and emotional arousal, without involving physical contact. However, this analysis was exploratory, and further research, including item development and confirmatory factor analysis, is required before immersive intimacy can be established as an independent construct.

## 4.5 Design Implications and Limitations

The present study showed that, in immersive video production, filming distance is not merely a staging choice, but a design variable that determines Presence and psychological closeness. In immersive video presented on a head-mounted display, camera position is experienced as the bodily position of the experiencer. Therefore, designing filming distance is not simply a matter of making the target appear larger; rather, it is an act of determining at what distance from the target the experiencer is positioned.

However, close-distance filming is not always optimal. As Proxemics research has shown, close distance may enhance intimacy and social involvement, whereas excessively close distance may induce discomfort associated with intrusion (Hall, 1966; Argyle & Dean, 1965; Probst, Rothe, & Hussmann, 2021). Thus, the appropriate filming distance in immersive video should be adjusted according to the content genre, the amount of performer movement, gaze direction and staging, the recipient’s prior involvement, and the expected experiential value. The present findings indicate that a design that does not leave experiencers merely as spectators of the content, but instead locates them near the target, may strengthen psychological closeness and the psychological basis of fan intensity.

The present study has at least five limitations. First, the experiment used specific idol content, and the findings may not generalise to all content genres. Second, the Avid fan and Casual fan groups differed in gender composition and age distribution. Although the main finding concerning the effect of Presence condition was based on within-participant comparisons and was not accompanied by an experience-order effect, interpretations involving prior-involvement group should be treated with caution. Third, although the study demonstrated a correspondence between Presence condition and psychological closeness, it did not identify which subcomponent of Presence most strongly explained the change in psychological closeness. Fourth, although the IOS scale is useful for measuring psychological closeness concisely, it is not a scale that measures fan intensity as a whole. Fifth, the study focused primarily on effects immediately after the experience, and the persistence of these effects was not sufficiently examined. Future research should use larger samples and test longitudinal models that include Presence, affective responses, psychological closeness, and its persistence.

# 5 CONCLUSION

In the present study, filming distance in 180° stereoscopic immersive video was manipulated to examine how the intensity of Presence corresponded to fan intensity towards the content target. The results showed that, compared with the low-Presence condition, the high-Presence condition significantly increased all Presence-related measures except the SUS index, with the largest condition differences observed for Possible Actions, Social Presence, and Observability. This confirmed that the filming-distance manipulation functioned as a manipulation that broadly enhanced Presence-related experience, particularly aspects closely tied to Place Illusion and the viewer’s situated relation to the performers. In addition, the analysis using IOS scale difference scores showed that the increase in fan intensity after the high-Presence video experience was significantly greater than that after the low-Presence video experience.

These findings indicate that filming distance in immersive video is not merely a matter of designing angle of view or composition, but a design variable that determines at what distance from the target experiencers locate themselves. Furthermore, enhanced Presence was shown to increase the extent to which the content target was experienced as close to the self, thereby functioning as a possible basis for deepening fan intensity.

## ACKNOWLEDGEMENTS

In conducting this experiment, the idol group STU48 performed the dance and vocals used as the experimental video stimuli. WATANABE Toru and IWAMI Shota (Concent, Inc.) assisted with the filming of the immersive video. NATSUKA Masato (CRAZY TV Creative, Inc.) assisted with the audio recording of the immersive video. The implementation of the immersive video for Apple Vision Pro was carried out by MATSUBARA Tatsuro (MESON, Inc.).

## CONTRIBUTIONS

KT, SMi, HH, YK, NY, and SMe conceived the research hypotheses. KT, SMi, HH, and SMe discussed the experimental design. KT, SMi, HH, NY, and SMe planned and conducted the filming of the experimental stimuli. KT, HH, NY, YK, and SMe conducted the experiment. KT, SMi, HH, NY, and SMe contributed to the interpretation and discussion of the findings. KT designed the experiment, analyzed the data, and wrote the manuscript. SMe was responsible for project administration, including the oversight and coordination of the broader Immersive Video research project. All authors contributed to the manuscript and approved the submitted version.

## CONFLICTS OF INTEREST

This study was conducted as a joint research project between MESON, Inc. and Hakuhodo DY Holdings Inc. With respect to the provision of research funding and the conduct of the study, there are no other conflicts of interest to disclose.

# SUPPLEMENTARY MATERIALS

## i Supplementary Information on Immersive Video

### i.i Filming distance and camera placement plan: (A) sectional view and (B) plan view.

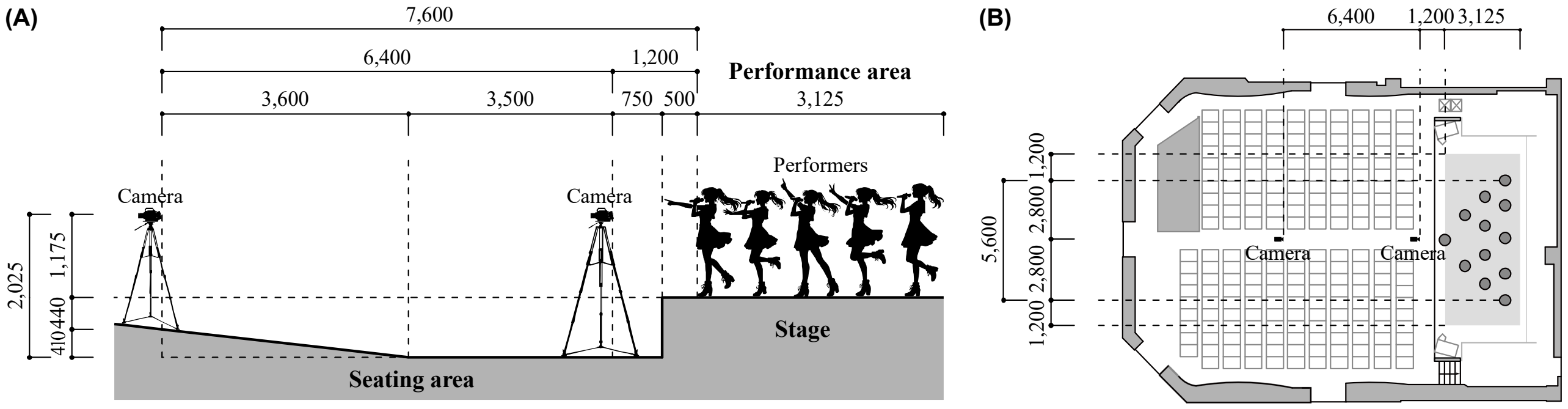

### i.ii Camera settings during filming and settings used during editing

| Stage | Parameter | Setting |
|---|---|---|
| Filming | Resolution per eye | 8,160 × 7,200 pixels |
| | Frame rate | 90 frames per second |
| | Shutter speed | 1/150 second |
| | Aperture value | F4.5, fixed |
| | ISO | 800 |
| | Focal length | 7.4 mm |
| | Neutral density filter | Not used |
| | White balance | Colour temperature: 5,600 Kelvin; tint: 10 |
| | Colour space | Blackmagic Design Film |
| | High dynamic range setting | High dynamic range |
| | Focus setting | Fixed focus, deep focus |
| | Compression ratio | 8:1 |
| | Lens calibration | Not performed during filming; calibrated before shipment |
| Editing | Colour grading | Manual adjustment and sharpness adjustment |
| | ISO | 400 |
| | Parallax adjustment | Not applied |
| | Resizing | 4,320 × 4,320 pixels per eye |

# ii Supplementary Information on the Exploratory Factor Analysis

## ii.i Scree plots: (A) Baseline, (B) post low-Presence condition, and (C) post high-Presence condition

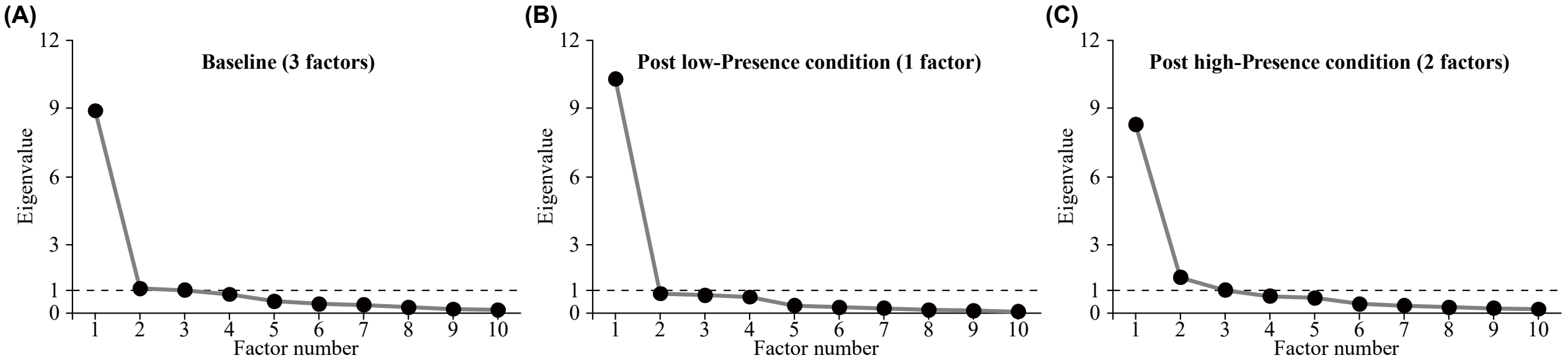

## ii.ii Factor-loading matrix (Note: Items are sorted based on the factor loadings in the post high-Presence condition)

| Negative term | Positive term | Baseline - 3 factors | | | | Post (LP) - 1 factor | | Post (HP) - 2 factors | | |
|---|---|---|---|---|---|---|---|---|---|---|
| | | Factor 1 | Factor 2 | Factor 3 | Primary F. | Factor 1 | Primary F. | Factor 1 | Factor 2 | Primary F. |
| Reluctant to support | Eager to support | 0.931 | — | — | Factor 1 | 0.892 | Factor 1 | 0.923 | — | Factor 1 |
| Not cute | Cute | 0.682 | — | — | Factor 1 | 0.846 | Factor 1 | 0.881 | — | Factor 1 |
| Gloomy | Cheerful | — | — | 0.877 | Factor 3 | 0.823 | Factor 1 | 0.772 | — | Factor 1 |
| Indifferent | Wishing for success | 0.901 | — | — | Factor 1 | 0.786 | Factor 1 | 0.752 | — | Factor 1 |
| Sad | Joyful | — | — | 0.706 | Factor 3 | 0.810 | Factor 1 | 0.709 | — | Factor 1 |
| Boring | Enjoyable | — | — | 0.419 | Factor 3 | 0.875 | Factor 1 | 0.701 | — | Factor 1 |
| Unpromising | Promising | — | — | — | — | 0.924 | Factor 1 | 0.638 | — | Factor 1 |
| Uninteresting | Interesting | — | 0.449 | 0.628 | Factor 3 | 0.787 | Factor 1 | 0.635 | — | Factor 1 |
| Uninspiring | Uplifting | 0.728 | — | — | Factor 1 | 0.941 | Factor 1 | 0.533 | 0.406 | Factor 1 |
| Unrelatable | Relatable | — | 0.673 | — | Factor 2 | 0.872 | Factor 1 | — | 0.849 | Factor 2 |
| Distant | Close | — | 0.998 | — | Factor 2 | 0.686 | Factor 1 | — | 0.832 | Factor 2 |
| Unmoved | Heart-racing | 0.462 | — | — | Factor 1 | 0.868 | Factor 1 | — | 0.734 | Factor 2 |
| Unexciting | Exciting | 0.497 | — | — | Factor 1 | 0.898 | Factor 1 | — | 0.564 | Factor 2 |
| Cold | Warm | — | — | — | — | 0.821 | Factor 1 | — | 0.521 | Factor 2 |

## ii.iii Dendrograms: (A) Baseline, (B) post low-Presence condition, and (C) post high-Presence condition

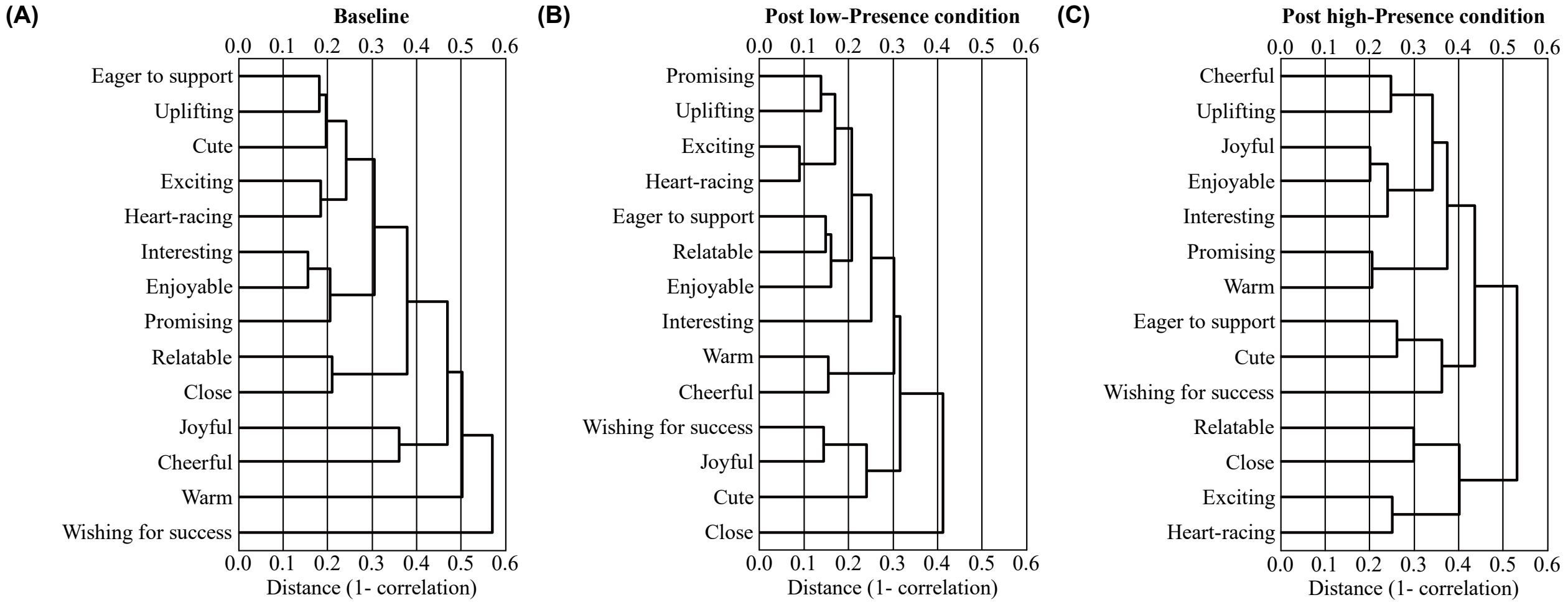

# iii Questionnaire Used in the Experiment

## iii.i Timing of questionnaire administration

| Section | Measure / Construct | Item no(s). | Baseline | Post1 | Post2 |
|---|---|---|---|---|---|
| 1 | Inclusion of Other in the Self Scale (IOS) | 01 | ✔ | ✔ | ✔ |
| 2 | Immediate Willingness / Hypothetical Willingness | 02 – 08 | ✔ | ✔ | ✔ |
| 3 | Current Affective State | 09 – 22 | ✔ | ✔ | ✔ |
| 4 | Presence – Core Presence | 23 – 27 | - | ✔ | ✔ |
| | Presence – Subdimensions | 28 – 40 | - | ✔ | ✔ |
| 5 | Behavioural Intentions: Willingness to Pay (WTP) | 41 – 48 | - | - | ✔ |
| 6 | Experience Quality (NPS / Live Attendance Intention / UGC / Satisfaction / Enjoyment / Rewatch Intention) | 49 – 58 | - | - | ✔ |
| 7 | Adverse Effects and Potential Confounds | 59 – 65 | - | - | ✔ |
| 8 | Comparative Evaluation / Content Evaluation | 66 – 72 | - | - | ✔ |
| 9 | Basic Demographics / Fan Activity Profile | 73 – 83 | - | - | ✔ |
| 10 | XR / Device Experience / Affinity for Technology Interaction (ATI-9) | 84 – 94 | - | - | ✔ |
| 11 | Future Interest and Use Scenarios | 95 – 98 | - | - | ✔ |

## iii.ii Full questionnaire items

| Administered at | Section | Construct / Subscale | Item no(s). | Item wording | Response format †1 |
|---|---|---|---|---|---|
| Baseline Post1 Post2 | 1 | IOS | 01 | Please select the figure that best represents the relationship between yourself and STU48. | [11-point IOS scale] Indicate the psychological closeness between yourself and STU48. |
| | 2 | Immediate Willingness | 02 | At this very moment, how much do you feel like learning more about STU48? | Not at all - Very much |
| | | | 03 | At this very moment, how much do you feel like viewing STU48 videos or listening to STU48 songs right after this? | |
| | | | 04 | At this very moment, how much do you feel like viewing STU48-related posts on social media or checking official accounts? | |
| | | | 05 | At this very moment, how much do you feel like talking about STU48 with a friend? | |
| | | Hypothetical Willingness | 06 | If a STU48 live performance were held tomorrow at a place you could attend, how much would you want to go? | Would not want to go - Would want to go |
| | | | 07 | If STU48 appeared on a programme or livestream tonight, would you want to adjust your other plans to watch it? | Would not want to watch - Would want to watch |
| | | | 08 | If tickets for a STU48 live performance were on sale now, would you want to purchase one? | Would not want to purchase - Would want to purchase |
| | 3 | Current Affective State | 09 | Boring – Enjoyable | Semantic differential scale |
| | | | 10 | Not cute – Cute | |
| | | | 11 | Uninteresting – Interesting | |
| | | | 12 | Uninspiring – Uplifting | |
| | | | 13 | Gloomy – Cheerful | |
| | | | 14 | Cold – Warm | |
| | | | 15 | Sad – Joyful | |
| | | | 16 | Distant – Close | |
| | | | 17 | Unrelatable – Relatable | |
| | | | 18 | Unmoved – Heart-racing | |
| | | | 19 | Unexciting – Exciting | |
| | | | 20 | Reluctant to support – Eager to support | |
| | | | 21 | Indifferent – Wishing for success | |
| | | | 22 | Unpromising – Promising | |
| Post1 Post2 | 4 | Core Presence Spatial Presence | 23 | To what extent did the video you experienced give you the feeling of being there? | Did not feel it at all - Felt it very strongly |
| | | Core Presence Slater-Usoh-Steed questionnaire (SUS) | 24 | During the experience, for how much of the time did the world inside the video become the reality that was dominant for you, to the extent that you forgot about the external real world? | Not at all - Almost all of the time |
| | | | 25 | When looking back on the video experience, was it closer to something you saw or to a place you visited? | Something I saw - A place I visited |
| | | Core Presence Self-Location | 26 | Did you feel that you were not merely observing, but were in the midst of the events? | Strongly disagree - Strongly agree |
| | | | 27 | To what extent did you feel that you were part of the video experience? | |
| | | Core Presence Social Presence | 28 | How often did you feel that you wanted to make eye contact with the members in the video? | Did not feel it at all - Felt it very strongly |
| | | | 29 | How often did you feel that the members in the video looked at you? | Never - Many times |

†1 Note: Unless otherwise stated, items were rated on a seven-point Likert scale.

| Administered at | Section | Construct / Subscale | Item no(s). | Item wording | Response format †1 |
|---|---|---|---|---|---|
| Post1<br>Post2 | 4 | Subdimensions<br>Possible Actions | 30 | During the experience, did you feel that you could touch the members in the video if you reached out your hand? | Did not feel it at all - Felt it very strongly |
| | | | 31 | When the members appeared to be moving towards you, how often did you feel that you wanted to move so as not to get in their way? | |
| | | Subdimensions<br>Perceptual Realism | 32 | Did you think the atmosphere and smell of the hall in the video would feel similar to those you would experience at an actual live venue? | Strongly disagree - Strongly agree |
| | | | 33 | Did you think the heat or coolness (temperature) of the hall in the video would feel similar to what you would experience at an actual live venue? | |
| | | Subdimensions<br>Interpersonal Distance<br>Intrusiveness | 34 | During the experience, how did you perceive the distance to the members? | Close - Far |
| | | | 35 | During the experience, did you feel that you wanted to move a little farther away from the STU48 members who were the subjects of the video? | Not at all - Very strongly |
| | | Subdimensions<br>Engagement | 36 | How often did you almost vocalise in response to the members? | Never - Many times |
| | | | 37 | How often did you smile in response to the members? | |
| | | | 38 | During this video, to what extent were you immersed without thinking about other things? | Not at all - Very much |
| | | Subdimensions<br>Observability | 39 | To what extent could you observe the members' facial expressions? | Could not observe them - Could observe them very clearly |
| | | | 40 | To what extent could you observe the members' clothing or costumes? | |
| Post2 | 5 | Behavioural<br>Intentions†2 | 41, 45 | Please select the amount closest to the maximum you would be willing to pay per session for an experience event where you could experience one song only. | [SA]<br>JPY 0<br>JPY 1-500<br>JPY 501-1,000<br>JPY 1,001-2,000<br>JPY 2,001-3,000<br>JPY 3,001-5,000<br>JPY 5,001-10,000<br>JPY 10,001 or more |
| | | | 42, 46 | If it were an experience event for one full performance (approximately 90 minutes), please select the amount closest to the maximum you would be willing to pay per session. | |
| | | | 43, 47 | Assuming that you owned an Apple Vision Pro, please select the amount closest to the maximum you would be willing to pay to purchase the content for one song. | |
| | | | 44, 48 | Please describe, to the extent you are comfortable, the rationale or motivation for the amount you selected above. | [Free text] |
| | 6 | Net Promoter Score<br>(NPS) | 49 | Regardless of video type, how likely would you be to recommend this experience event (180-degree video experience event) to a friend or acquaintance? | [11-point scale]<br>Would not recommend at all - Would definitely recommend |
| | | Live Attendance<br>Intention | 50 | After this experience event, if STU48 were to hold a live performance on a date and at a location you could attend, would you want to go? | Strongly disagree - Strongly agree |
| | | User-Generated<br>Content Intention<br>(UGC) | 51 | Regardless of video type, would you want to post or share about this experience on social media, etc.? | |
| | | | 52 | Regardless of video type, I would want to tell others about this experience on social media, etc. | |
| | | Satisfaction<br>Enjoyment<br>Rewatch Intention†2 | 53, 56 | I was satisfied with the overall experience. | |
| | | | 54, 57 | This experience was enjoyable. | |
| | | | 55, 58 | I would like to experience a similar video again. | |
| | 7 | Cybersickness<br>(shortened SSQ) | 59 | During or immediately after the experience, I felt nausea. | [5-point scale]<br>1 = not at all<br>5 = symptom present |
| | | | 60 | During or immediately after the experience, I felt dizziness or unsteadiness. | |
| | | | 61 | During or immediately after the experience, I felt eye fatigue or eyestrain. | |
| | | | 62 | During or immediately after the experience, I felt a headache. | |
| | | Experience<br>Disruptors | 63 | Image quality, sound quality, latency, or related factors bothered me and made it difficult to concentrate on the experience. | [5-point scale]<br>1 = did not feel it at all<br>5 = felt it very strongly |
| | | | 64 | The fit or feel of the Apple Vision Pro (weight, pressure, heat, etc.) bothered me and made it difficult to concentrate on the experience. | |
| | | | 65 | The laboratory environment (noise, temperature, etc.) bothered me and made it difficult to concentrate on the experience. | |
| | 8 | Comparative<br>Evaluation of Video<br>Conditions | 66 | Which did you prefer, Video A or Video B? | [SA]<br>Video A (High-Presence)<br>Video B (Low-Presence) |
| | | | 67 | Please explain why you selected Video A or Video B. | [Free text] |
| | | Favourite Member<br>Scene Evaluation | 68 | Were any of your favourite members included in the videos you experienced? | [MA]<br>STU48 as a whole<br>STU48 fourth-generation trainees as a whole<br>One of the 12 members<br>None of the above (another fourth-generation trainee)<br>None of the above (another STU48 member)<br>None |

†2 Note: For paired item numbers, the same item was administered separately for the high-Presence and low-Presence videos.

| Administered at | Section | Construct / Subscale | Item no(s). | Item wording | Response format †1 |
|---|---|---|---|---|---|
| Post2 | 8 | Favourite Member Scene Evaluation | 69 | Did you find a new favourite member through this video experience? | [SA] Yes / No |
| | | | 70 | If there were any scenes you liked in this video experience, please select them. | [MA]<br>Scenes where members were singing<br>Scenes where you made eye contact with members<br>Scenes where members gave you a response<br>Scenes where all 12 members stood side by side<br>Scenes where you could see the whole stage within the field of view<br>Scenes where sound was heard spatially<br>Scenes where footstep sounds from members moving were included<br>Other |
| | | Distance Evaluation of the Close-Distance Condition | 71 | How did you perceive the closeness to the members in Video A? | [SA]<br>Too close<br>Somewhat too close<br>Just right<br>Could be slightly farther<br>Could be much farther |
| | | Open-Ended Evaluation of the Content | 72 | Please freely write your impressions of the content you experienced | [Free text] |
| | 9 | Basic Demographics | 73 | Please indicate your gender. | [SA] Male / Female |
| | | | 74 | Please indicate your age. | [Free text] |
| | | | 75 | Please indicate your industry. | [SA]<br>Choices based on the classification by the Ministry of Internal Affairs and Communications |
| | | | 76 | Please indicate your occupation or job type. | |
| | | Fan Activity Profile | 77 | Are you a member of the STU48 fan club? | [SA]<br>Member<br>Non-member |
| | | | 78 | Please select all STU48 social media accounts that you follow. | [MA]<br>STU48 official X<br>STU48 members' X<br>STU48 fourth-generation trainees official X<br>STU48 official LINE<br>STU48 official Instagram<br>STU48 members Instagram<br>STU48 fourth-generation trainees official Instagram<br>YouTube<br>TikTok<br>None |
| | | | 79 | On how many days in the past week did you check STU48-related information? | [SA]<br>0 days<br>1-2 days<br>3-4 days<br>5-6 days<br>Every day |
| | | | 80 | How many times in the past week did you watch STU48-related videos or livestreams? | [SA]<br>0 times<br>1-2 times<br>3-4 times<br>5-6 times<br>Every day |
| | | | 81 | In the past three months, how many times have you attended STU48-related in-person events (theatre performances, live performances, events, etc.)? | [SA]<br>0<br>1<br>2<br>3-4<br>5 or more |
| | | | 82 | Approximately how much have you spent on STU48 in the past six months? | [SA]<br>JPY 0<br>Up to JPY 5,000<br>Up to JPY 10,000<br>Up to JPY 20,000<br>Up to JPY 50,000<br>Up to JPY 100,000<br>Over JPY 100,000 |
| | | | 83 | Which of the following best describes your style of oshi-katsu or your way of being a fan? | [SA]<br>six Oshinomics categories †3 |

†3 Note: Oshinomics is a fan-classification framework developed from a large-scale survey of idol fans, in which fans are categorised into six clusters (a survey conducted by Hakuhodo).

https://www.hakuhodo.co.jp/humanomics-studio/assets/pdf/OSHINOMICS_Report.pdf

| Administered at | Section | Construct / Subscale | Item no(s). | Item wording | Response format †1 |
|---|---|---|---|---|---|
| Post2 | 10 | XR / Device Experience | 84 | Please indicate your experience with Apple Vision Pro. | [SA]<br>Own and use Apple Vision Pro<br>Use Apple Vision Pro for work, etc.<br>Have experienced an Apple Vision Pro demonstration<br>Experienced Apple Vision Pro for the first time today |
| | | | 85 | Please indicate your experience with head-mounted displays (HMDs) other than Apple Vision Pro. | [SA]<br>Own and use an HMD such as Quest<br>Use an HMD for work, etc.<br>Have used an HMD<br>Have never worn an HMD |
| | | Affinity for Technology Interaction (ATI-9) | 86 | I like to engage deeply with technical systems and explore them in detail. | [6-point scale]<br>Completely disagree<br>Strongly disagree<br>Slightly disagree<br>Slightly agree<br>Strongly agree<br>Completely agree |
| | | | 87 | I like to try out the functions of new technical systems. | |
| | | | 88 | I often deal with technical systems because it is necessary. | |
| | | | 89 | When I encounter a new technical system, I intensively try out various things. | |
| | | | 90 | I enjoy taking time to become familiar with new technical systems. | |
| | | | 91 | As long as it works, that is enough; I am not interested in how or why it works. | |
| | | | 92 | I try to understand exactly how technical systems operate. | |
| | | | 93 | It is sufficient for me to understand only the basic functions of technical systems. | |
| | | | 94 | I try to make the fullest possible use of the capabilities of technical systems. | |
| | 11 | Future Interest and Use Scenarios | 95 | What genres of works or videos would you like to experience in a 180-degree video format like this one? | [MA]<br>Making-of footage<br>Documentaries<br>Live performance footage<br>Drama/film<br>Music videos<br>Animation/3D CG<br>Musicals/Opera<br>Sports/Matches<br>Travel/Tourism<br>Exhibitions/Museums, etc.<br>Traditional performing arts such as Kabuki or Noh<br>Traditional crafts such as ceramics or swords<br>History/Historical events<br>Other |
| | | | 96 | What types of people or content subjects would you like to experience in a 180-degree video format like this one? | [MA]<br>Idols<br>Actors/Actresses<br>Singers/Musicians<br>Athletes<br>Dancers/Performers<br>Cultural figures/Craftspeople in traditional performing arts and related fields<br>Models/Fashion<br>Artists such as painters or photographers<br>Other |
| | | | 97 | If you were to experience something like this together with an acquaintance, what situation would you prefer? | [MA]<br>Each of you wears an Apple Vision Pro and experiences it while talking in the same room<br>Each of you wears an Apple Vision Pro and experiences it while talking from remote locations via the internet<br>You and a friend enter a virtual space as avatars and experience it as if viewing side by side at a live venue<br>Instead of wearing a device such as Apple Vision Pro, you view it side by side on a large cinema-like screen |
| | | | 98 | Please freely write your impressions of the overall experience event. | [Free text] |